
\input harvmac
\input epsf

%
%
%

\def\tilde{\widetilde}
\def\bar{\overline}
\def\hat{\widehat}
\def\*{\star}
\def\[{\left[}
\def\]{\right]}
\def\({\left(}		
\def\){\right)}

%
%
\def\zb{{\bar{z} }}
\def\frac#1#2{{#1 \over #2}}
\def\inv#1{{1 \over #1}}

\def\d{\partial}
\def\der#1{{\partial \over \partial #1}}

\def\2pi{\hbox{$2\pi i$}}

\def\dsl{\raise.15ex\hbox{/}\kern-.57em\partial}
\def\Dsl{\,\raise.15ex\hbox{/}\mkern-.13.5mu D}
%
%
\def\th{\theta}		\def\Th{\Theta}
		
\def\be{\beta}
\def\al{\alpha}
\def\ep{\epsilon}
\def\la{\lambda}	
		\def\De{\Delta}

%
%
		\def\CC{{\cal C}}
		
	\def\CH{{\cal H}}

\def\CP{{\cal P}}		
	\def\CT{{\cal T}}	\def\CU{{\cal U}}

\def\2pi{\hbox{$2\pi i$}}

\def\dsl{\raise.15ex\hbox{/}\kern-.57em\partial}
\def\Dsl{\,\raise.15ex\hbox{/}\mkern-.13.5mu D}
%
%
%
\font\numbers=cmss12
\font\upright=cmu10 scaled\magstep1
\def\stroke{\vrule height8pt width0.4pt depth-0.1pt}
\def\topfleck{\vrule height8pt width0.5pt depth-5.9pt}
\def\botfleck{\vrule height2pt width0.5pt depth0.1pt}
\def\Zmath{\vcenter{\hbox{\numbers\rlap{\rlap{Z}\kern
0.8pt\topfleck}\kern
2.2pt
                   \rlap Z\kern 6pt\botfleck\kern 1pt}}}
\def\Qmath{\vcenter{\hbox{\upright\rlap{\rlap{Q}\kern
                   3.8pt\stroke}\phantom{Q}}}}
\def\Nmath{\vcenter{\hbox{\upright\rlap{I}\kern 1.7pt N}}}
\def\Cmath{\vcenter{\hbox{\upright\rlap{\rlap{C}\kern
                   3.8pt\stroke}\phantom{C}}}}
\def\Rmath{\vcenter{\hbox{\upright\rlap{I}\kern 1.7pt R}}}
\def\Z{\ifmmode\Zmath\else$\Zmath$\fi}
\def\Q{\ifmmode\Qmath\else$\Qmath$\fi}
\def\N{\ifmmode\Nmath\else$\Nmath$\fi}
\def\C{\ifmmode\Cmath\else$\Cmath$\fi}
\def\R{\ifmmode\Rmath\else$\Rmath$\fi}

\Title{CLNS 92/1150, HUTP-92/A045 }
{\vbox{\centerline{Quantum Affine Symmetry as  }
\centerline{ Generalized  Supersymmetry} }}

\bigskip
\bigskip

\centerline{Andr\'e LeClair}
\medskip\centerline{Newman Laboratory}
\centerline{Cornell University}
\centerline{Ithaca, NY  14853}
\bigskip\bigskip

\centerline{Cumrun Vafa}
\medskip\centerline{Lyman Laboratory of Physics}
\centerline{Harvard University}
\centerline{Cambridge, MA 02138}

\vskip .3in

The quantum affine $\CU_q ( \hat{sl(2)} ) $ symmetry
is studied when $q^2$ is an even root of unity.
The structure of this algebra allows  a natural
generalization of $N=2$ supersymmetry algebra.  In particular
it is found that the momentum
operators $P ,\bar{P}$, and thus the Hamiltonian,
can be written as generalized multi-commutators,
and can be viewed as new central elements of the algebra
$\CU_q (\hat{sl(2)} )$.
We show that massive particles in (deformations of)
integer spin representions
of $sl(2)$
are not allowed in such theories.  Generalizations of Witten's index
and
Bogomolnyi bounds are presented and a preliminary
attempt in constructing manifestly $\CU_q (\hat{sl(2)} )$ invariant
actions
as generalized supersymmetric Landau-Ginzburg theories is made.

\Date{9/92}
%
%
%
%
%
%
\def\ro{{\hat{\rho}^{(1/2)}}}
\def\uqo{{ \CU^{(0)}_q }}
\def\dag{{\dagger}}
\noblackbox
\def\rh{{\hat{R}}}
\def\ot{\otimes}
\def\phib{\bar{\phi}}
\def\qb#1{{\bar{Q}_{#1}} }
\def\q#1{ { Q_{#1} } }
\def\zb{{\bar{z}}}
\def\uq{\CU_q}
\def\a{{\, {\rm ad} \, }}
\def\pb{\bar{P}}
\def\Qb{{\bar{Q}}}
\def\qp{{Q_+}}
\def\qm{{Q_-}}

\newsec{Introduction }

One of the early discoveries in particle physics was the
realization of the existence of two different types of particles:
Bosons and
Fermions.  In the algebraic context they are distinguished by
the fact that bosonic operators generally satisfy simple
commutation relations whereas fermionic operators satisfy simple
anti-commutation relations.  In quantum field theories which
are  supersymmetric, there is a strong connection between
these two classes of operators. Supersymmetry gives a simple
organizing principle for
a theory to have both types of these particles in a symmetrical
fashion.

In the context of 2 dimensional quantum field theories
it is natural
to expect that one would encounter more exotic objects
than just bosons and fermions.  For example,  the
experience with rational conformal field theories suggests
that the braiding properties of non-local operators is a
key ingredient in any attempt at
understanding the structure of such theories.  In these cases
the braiding properties are for fields  with
`fractional spin'.
In 2 dimensions, spin is defined with respect to Lorentz boosts,
or equivalently, Euclidean rotations.  Fields can have definite
spin, however since the Lorentz boost generator $L$ does not commute
with the Hamiltonian $H$, asymptotic particles have no definite
spin.

{\it Quantum
groups}
\ref\drin{V. G. Drinfel'd, Sov. Math. Dokl. 32 (1985) 254; Sov. Math.
Dokl.
36 (1988) 212.}\ref\jimbo{M. Jimbo, Lett. Math. Phys. 10 (1985) 63;
Lett. Math. Phys. 11 (1986) 247.}\
can be formulated such that the defining
relations  are expressed as  generalized
commutation relations which include phases in
the quadratic relations instead of just plus or minus signs.
In this way, quantum groups present themselves as natural candidates
for generalizations of supersymmetry.
Indeed it has been discovered that many integrable 2d QFT's,
such as the  sine-Gordon theory and its generalizations to the
ordinary  supersymmetric and fractional supersymmetric sine-Gordon
theories,
and also the generalization of all of these to other affine Toda
theories,
enjoy a quantum affine symmetry
\ref\rbl{D. Bernard and A. LeClair, Phys. Lett. B247
(1990) 309; Commun. Math. Phys. 142 (1991) 99.}.
Furthermore, at  special points the quantum affine
symmetry is nothing other than $N=2$ supersymmetry.

It is natural to ask how much of the
structure of supersymmetric theories carries over to the
context of quantum affine symmetry.
This is an important question to settle in view
of the fact that two dimensional quantum field thoeries
with $N=2$ supersymmetry have been shown to have
a very rich structure
\ref\cv{S. Cecotti and C. Vafa, Nucl. Phys. B367 (1991)
359.}\ref\newind{S.
Cecotti, P. Fendley, K. Intrilligator, and C. Vafa, `A New
Supersymmetric Index',  preprint HUTP-92/A021, SISSA 68/92/EP,
BUHEP-92-14.}.
The $N=2$ superalgebra has important features
that have not been explored before in the context of the
algebraic properties of quantum affine algebras.
These include the existence of chiral rings
\ref\lvw{W. Lerche,C. Vafa and N.P. Warner, Nucl. Phys. B324 (1989)
427 .}\  and  of a topological twisting
\ref\EW{E. Witten, Comm. Math. Phys. 118 (1988) 411.}.
  Our paper is a first step toward understanding
these novel properties of quantum affine algebras.
In particular we show that many of the
properties
 of the $N=2$ supersymmetry algebra are present
in the
quantum affine $\CU_q (\hat{sl(2)} )$ algebra  with the quantum
deformation
parameter $q^2$
satisfying $q^{2p}=1$, where $p$ is an even integer.  In particular
we
have  four (fractional spin) supercharges
$Q_\pm , \Qb_\pm $ which are nil-potent $Q_\pm^p= \Qb_\pm^p = 0$,
which
generalize the  supersymmetric ($p=2$) result.
More importantly, introducing the light-cone components $P,~ \pb$ of
the momentum vector operator, where the Hamiltonian $H= P + \pb$,
we find that $P, ~\pb$
can be expressed in terms of the quantum affine generators, again
generalizing the $N=2$
supersymmetric result  $P=\{ Q_+, Q_-\}$; the new subtlety
is that $P$ is no longer quadratic in $Q$'s.  Rather $P$ is
a polynomial in $Q$'s each term of which has equal number $(p/2)$ of
$Q_+$ and $Q_-$'s.  Similar results apply to $\pb$ with the
charges $\Qb_\pm$, such that the Hamiltionian is in the universal
enveloping algebra of the quantum affine generators.
As we will explain, these same elements can also correspond
to other local integrals of motion with higher odd integer spin.

The main results of our paper have
a purely algebraic characterization.
We show that the  operators $P , \pb$
are  new central elements in the quantum
affine $\CU_q (\hat{sl(2)})$ algebra
with quantum parameter an even root of unity,
which have trivial comultiplication and are `neutral'
(are in the `purely imaginary direction' in the mathematical
description).
As we will show, the explicit form of the elements $P,\pb$
is completely characterized by the properties of the algebra
$\CU_q (\hat{sl(2)} )$.

The organization of the paper is as follows:  In section 2 we
formulate
the problem and provide examples ($p=2,4$) of solutions
within  the purely algebraic framework of the algebra
$\CU_q (\hat{sl(2)} )$.
In section 3 we
formulate the problem generally in the context of  quantum field
theory.
The main result of this
section is a condition for constructing  local conserved quantities
out of generalized multi-commutators of the non-local $Q_\pm$'s. The
latter condition
is expressed in terms of the braiding matrices of the non-local
currents.
In section 4 we present the solutions to these conditions
for some higher examples $(p=6,8)$.  In section
5 we illustrate  how these ideas are realized in a
concrete example (sine-Gordon (SG) theory
at  special values of the coupling).
In section 6 we describe a first attempt at applying these new
ideas.  In particular we find the analog of the Bogomolnyi mass bound
for kinks, and the absence of particles with integral $sl(2)$ spin.
We also discuss some aspects of a superspace Landau-Ginzburg
formulation for theories with affine quantum symmetry.
In section 7 we present our conclusions and suggest
some directions for future investigation.
In appendix A, some low dimensional representations of the quantum affine
algebra are given. In appendix B, some spectral properties
of representations (related to the absence of physically interesting
integral spin representations) are proven.

\newsec{Algebraic Formulation}

\medskip\noindent
2.1 ~~{\it The $\CU_q (\hat{sl(2)} )$ Quantum Affine Algebra}
\medskip

In this section we describe how to formulate our results purely
algebraically.
The $\uq \equiv \CU_q (\hat{sl(2)} ) $
loop algebra\drin\jimbo\ is the universal enveloping algebra
generated by $\q \pm ,\qb \pm $, and $T$, satisfying\foot{The
presentation we use for $\uq$ is not standard, but is more
suitable for field theory applications. The more
standard presentation of the algebra $\uq$ is with respect to the
basis
$\{ e_i ,f_i ,h_i ,
i=0,1\}$ with generalized Cartan matrix $\{a_{ij} \} =
\left(\matrix{2&-2\cr -2 & 2\cr}\right)$.  Up to constants the
isomorphism is
$\q+ = e_1 q^{h_1 /2} $, $\q- = e_0 q^{h_0 /2} $, $\qb+ = f_0 q^{h_0
/2} $,
$\qb- = f_1 q^{h_1 /2}$, $h_1 =T, h_0 = -T + k$, where $k$ is the
central extension (which corresponds to the level in the undeformed
affine Lie algebra).
We have set $k=0$ since this corresponds to the known physical
applications.
}
\eqn\IIi{\eqalign{
[T, \q\pm ] = \pm 2 \q\pm , ~~~&~~~ [T, \qb\pm ] = \pm 2 \qb\pm
 \cr
\q\pm \qb\pm - q^2 \qb\pm \q\pm &= 0
\cr
\q\pm \qb\mp - q^{-2} \> \qb\mp \q\pm  &=  a^2 \frac{\( 1- q^{\pm 2T}
\)}
{(1-q^2)} ,
\cr }}
where $a^2$ is some arbitrary constant, which we take to be positive.
In a physical realization, $a$ is usually dimensionful.
The algebra $\uq$ is a q-deformation of the affine Lie algebra $\hat{sl(2)}$,
the above generators corresponding to the Cartan-Weyl basis of simple roots.
The
ordinary $N=2$ supersymmetry algebra corresponds to $q^2=-1$
(where $a\not= 0$ in the last commutation relation {\it only} in
infinite
volume where central terms may exist).
Notably absent from the above commutation relations is the
commutation properties between $Q_+$ and $Q_-$ (and the right-moving
counterparts).  In the quantum affine algebras there is no
simple commutation properties for these operators.   Instead
they are replaced by the deformed Serre relations,
which will play an important role in the sequel:
\eqn\IIii{\eqalign{
Q^3_\pm \q \mp  - (1+q^2+q^{-2}) Q^2_\pm \q\mp \q\pm
+ (1+q^2 + q^{-2}) \q\pm \q\mp Q^2_\pm - \q\mp Q^3_\pm &= 0\cr
\Qb^3_\pm \qb \mp  - (1+q^2+q^{-2}) \Qb^2_\pm \qb\mp \qb\pm
+ (1+q^2 + q^{-2}) \qb\pm \qb\mp \Qb^2_\pm - \qb\mp \Qb^3_\pm &= 0
. \cr }}
The motivation for imposing these relations is discussed below.
For the $N=2$ supersymmetric theories the Serre relations
are satisfied in a trivial way as $Q^2=0$ for each supercharge.
Moreover the anti-commutation of $Q_+$ and $Q_-$ give $P$
(and their right-moving counterparts).  It is precisely the
apparent absence of such relations in the general quantum affine
algebra that motivated us to find them.

The algebra $\uq$ is a Hopf algebra with comultiplication
$\De$, counit $\ep$, and antipode $S$:
\eqn\IIiii{\eqalign{
\De (T) &= T \ot 1 + 1\ot T \cr
\De (\q\pm ) &= \q\pm \ot 1 + q^{\pm T} \ot \q\pm \cr
\De (\qb\pm ) &= \qb\pm \ot 1 + q^{\mp T} \ot \qb\pm  \cr
\ep (\q\pm ) &= \ep (\qb\pm) = \ep(T) = 0 \cr
S(T) &= -T ,~~~~~ S(\q\pm) = -q^{\mp T} \> \q\pm , ~~~~~
S(\qb\pm) = - q^{\pm T} \> \qb\pm
. \cr}}
These definitions satisfy the usual properties of a Hopf algebra:
\eqn\IIiiib{\eqalign{
\De (A) \De(B) &= \De(AB) \cr
S(AB) &= S(B) S(A) \cr
\ep ( AB) &= \ep(A) \ep (B) \cr
m (S\ot id ) \De (A) &= m (id \ot S ) \De (A) = \ep (A) \cr
(\ep \ot id ) \De &= (id\ot \ep ) \De = id , \cr }}
where $A,B \in \uq$, and $m$ is the multiplication map.
The first equation defines $\De$ to be a homomorphism
from $\uq \to \uq\ot \uq$.

In a field theory realization of $\uq$ as a symmetry algebra,
these Hopf algebra properties have a precise meaning\foot{
See
\ref\gs{C. Gomez and G. Sierra, Nucl.
Phys.
B 352 (1991) 791.}\ref\double{D. Bernard
and A. LeClair,
`The Quantum Double in Integrable Quantum Field Theory', preprint
CLNS 92/1147, SPhT-92-054,
to appear in Nucl. Phys. B.}\
for a detailed discussion. Some of this is reviewed in  section 3.}.
For the discussion in this section we only need to recall that
the comultiplication $\De$ describes how the elements of
$\uq$ act on products of fields at different space-time locations,
or how they act on multiparticle states.
The antipode $S$ corresponds to a Euclidean rotation by an angle
$\pi$ in the $x-t$ plane, and the counit $\ep$ is the 1-dimensional
vacuum representation of $\uq$.

Because $\uq$ is a deformation of an affine Lie algebra, there
exists a meaningful derivation $d$, which in the principal
gradation\foot{This
gradation is a twist (inner automorphism) of the more familiar
gradation
(the homogeneous one)
encountered in the affine structure of current algebra. We emphasize
that
the physical properties (such as Lorentz spin)
are not invariant under these changes of gradation,
and it is the principal gradation that is relevent in e.g. the SG
model. The homogeneous gradation becomes relevent in the twisted
version (see sections 6.3, 6.4).}
is defined to satisfy
\eqn\IIx{
[d, T] = 0,~~~~~[d, \q\pm ] = \q\pm , ~~~~~[d,\qb\pm ] = - \qb\pm . }
In the field theory, $d$ is proportional to the Lorentz boost
generator $L$.  Namely, if $q^2 = \exp (-2\pi i \al )$,
then $L= s\, d$, where $s=\al ~~{\rm Mod} (\Zmath)$\rbl. (See section
3.)
The equation \IIx\ implies that the Lorentz spin of the generators
$Q_\pm$
 ($\Qb_\pm$) is thus $s$ ($-s$).
We take $s=\al$.

We introduce the basic
two dimensional representation $\ro $ of $\uq$ on the vector
space $V \sim \Cmath^2$:
\eqn\IIxxii{
\ro (T) = \sigma_3,~~~~~\ro (\q\pm ) = \frac{a}{2} \, \nu \,
\sigma_\pm ,
{}~~~~~\ro (\qb\pm ) = \frac{a}{2} \, \nu^{-1} \,    \sigma_\pm ,
{}~~~~~\nu = e^{s \, \th } , }
where
$\sigma_i$ are the
Pauli spin matrices\foot{
$\sigma_3 = \left(\matrix{1&0\cr 0 & -1 \cr } \right)$,
$\sigma_+ = \left(\matrix{0&2\cr 0 & 0 \cr } \right)$,
$\sigma_- = \left(\matrix{0&0\cr 2 & 0 \cr } \right)$.
}.
In anticipation of the field theory realization, we have
set the spectral or `loop' parameter $\nu$  to $\exp ( s\,\th  )$.
(In e.g. the SG theory $V$ is the vector space of one-soliton states
and $\th$  is the rapidity of on-shell particles.)
A finite  Lorentz boost generated by $\exp (\al \, L )$
shifts $\th \to \th + \al$, thus the above representation
satisfies \IIx.

We introduce the following Hermitian conjugation properties:
\eqn\IIiv{
T^\dag = T,~~~~~Q^\dag_\pm = \q\mp ,~~~~~\Qb^\dag_\pm = \Qb_\mp
,~~~~~
q^\dag = q^{-1} }
which are compatible with the relations \IIi\IIii\ when $q$ is a
phase.
In the SG theory these Hermiticity properties can be deduced from the
explicit field theory expressions.  We remark that this Hermiticity
structure is not the conventional one (encountered in
e.g. current algebra), and is only possible because
we are in the principal gradation.

For an arbitrary Hopf algebra $\CU$, one defines adjoint actions as
follows.
Let
\eqn\IIv{
\De (A) = \sum_i a_i \ot b_i ,}
with
$A,a_i,b_i \in \CU$, and define
\eqn\IIvi{
\a A (B) = \sum_i a_i \> B \> S(b_i ) . }
This adjoint action satisfies the following important properties:
\eqn\IIvii{\eqalign{
\a A \> \a B (C) &= \a \>AB\> (C) \cr
\a A (BC) &= \sum_i \a a_i (B) \> \a b_i (C) . \cr } }
The first relation in \IIvii\ is equivalent to saying that viewed
as a map $\CU \ot \CU \to \CU$, $\a$ is a $\CU$ homomorphism.
Thus the adjoint action generally induces representations of $\CU$.
For ordinary Lie algebras, $\a A\, (B) = [A,B]$, the first relation
in
$\IIvii$ implies the Jacobi relations, and the second relation is the
statement $[A, BC] = [A,B] C + B [A,C]$.

For the algebra $\uq$, the adjoint actions take the form
\eqn\IIviii{\eqalign{
\a T (A) & = [T,A]~, ~~~~~~~~~~
\a T (AB) = \a T (A) B + A \a T (B) \cr
\a \q\pm (A) &= \q\pm A -  q^{\pm \a T  } (A) \>  \q\pm  \cr
\a \qb\pm (A) &= \qb\pm A -  q^{\mp \a T} (A) \>  \qb\pm \cr
\a \q\pm (AB) &= \a \q\pm (A) B  + q^{\pm  \a T} (A) \a \q\pm (B) \cr
\a \qb\pm (AB) &= \a \qb\pm (A) B  + q^{\mp  \a T } (A) \a \q\pm (B)
, \cr}}
where
$$q^{\pm \a T } (A) \equiv q^{\pm T} \, A \, q^{\mp T} = q^{\pm T_A}
\> A.$$
($T_A$ is defined by the equation $[T,A] = T_A \, A$.)
The relations \IIi\ and \IIii\ may be expressed using these $\a$'s:
\eqn\IIix{\eqalign{
\a T (\q\pm )  &= \pm 2 \q\pm , ~~~~~ \a T (\qb\pm )  = \pm 2 \qb\pm
 \cr
\a \q\pm (\qb\pm ) &= 0
\cr
\a \q\pm (\qb\mp )  &=  \frac{\( 1- q^{\pm 2T} \)}
{(1-q^2)} ,
\cr }}
\eqn\IIixb{
\a^3 \q\pm (\q\mp ) = \a^3 \qb\pm (\qb\mp ) = 0. }

The Serre relations have the following meaning.  Let $\CU^{(0)}_q$
denote
the finite $\CU_q (sl(2))$ subalgebra generated by ${\q+,\qb-,T}$, and let
$\a \CU^{(0)}_q $ denote $\a A$ for any $A\in \CU^{(0)}_q$.
Since $\a \uqo$ is a $\uqo$ homomorphism, $\a \uqo (Q_- )$ induces
a representation of $\uqo$.   The relations
$\a^3 \q+ (\q-) = \a \qb- (\q- ) = 0$ mean that $\a \uqo$ on
$\q-$ induces a 3 dimensional representation.  As $q\to 1$, the
latter representation is of course isomorphic to the adjoint
representation
of $sl(2)$.

\bigskip
\noindent
2.2 ~~{\it Algebraic construction of Local Integrals of Motion}

\medskip

We will now formulate in purely algebraic terms what one
needs in order to construct new local integrals of motion
out of $Q$'s.
Let us suppose that the elements of $\uq$ are conserved charges in
some
Lorentz invariant 2 dimensional physical system.  Introduce the
Euclidean space-time translation, or momentum, operators $P,\pb$.
They act on fields $\Psi (z,\zb )$ as follows:
\eqn\IIxi{
\[ P, \Psi (z,\zb ) \] = \d_z \Psi (z,\zb ) , ~~~~~~
\[ \pb, \Psi (z,\zb ) \] = \d_\zb \Psi (z,\zb ) , }
where
$z = (t+ix)/2 , \zb = (t-ix)/2 $.  The Hamiltonian of the
system is $H=P + \pb$, whereas $P_x = P - \pb$ is the
spacial translation operator.

Let us further hypothesize that it is possible to express
$P,\pb$ as elements in $\uq$.  Then $P,\pb$ are
subject to the following purely algebraic conditions:

\noindent
(i) ~~$P$ and $\pb$ are self-Hermitian conjugate:
\eqn\IIxii{
P^\dag = P , ~~~~~~~ \pb^\dag = \pb . }

\noindent
(ii) ~~ The operators $P,\pb$ have Lorentz spin $\pm 1$:
\eqn\IIxiii{
[L, P ] = P , ~~~~~~~ [ L, \pb ] = - \pb . }

\noindent
(iii) ~~ $P, \pb$ are central elements of $\uq$:
\eqn\IIxiv{
\a A \, (P) = \a A \, (\pb ) = 0 ~~~~~~~~~\forall ~ A \in \uq  .}
This is simply the statement that the elements of $\uq$ are
conserved since $\d_t A = [H,A]$.
The centrality of $P,\pb$ implies also that $[P,\pb ] = 0$;
the latter relation together with \IIxiii\ is the Poincar\'e
algebra in two dimensions.

\noindent
(iv) ~~ $P,\pb$ must have the trivial comultiplication:
\eqn\IIxv{
\De (P) = P\ot 1 + 1 \ot P ,}
and similarly for $\pb$. This is simply the statement that
$P,\pb$ are local conserved quantities.  This implies for instance
that the energy-momentum of a multiparticle state is the sum of the
individual energy-momentum.  It is also evident from the action of
$P$ on a product of two fields:
$$ \[ P, \Psi \> \Psi' \] = (\d_z \Psi) \> \Psi ' + \Psi \> ( \d_z
\Psi ') .$$

The above conditions, though easily formulated on physical grounds,
are rather non-trivial algebraically.  Indeed, for generic
$q$ they are impossible to satisfy for numerous reasons. If
$q$ is generic, the Lorentz spin $\pm s$ of the charges $\q\pm ,
\qb\pm$ is irrational, thus one cannot construct operators
$P,\pb$ out of them with spin $\pm 1$.  Suppose now that
$q$ is the following root of unity:
\eqn\IIxvi{
q^2 = \exp (-2\pi i / p ) ,   ~~~~~\Rightarrow s = 1/p , }
where $p$ is a positive even integer.
One is led to the following minimal ansatz for a solution
to the conditions (i)-(iv).  Define $\uq^+ \subset \uq$ to be
generated by $\q\pm$, and similarly $\uq^- \subset \uq$ to be
generated
by $\qb\pm$.  We restrict our attention to finding solutions of
(i)-(iv)
of the following type:
\eqn\IIxvii{
P\in \uq^+,~~~~~\pb \in \uq^- . }
The elements $P, \pb$ we will construct
are actually in a subalgebra of $\uq^\pm$ consisting of the
elements obtained from the basic generators via adjoint action.
Consider first the element $P$. The Lorentz spin condition (ii),
together with
the constraint that $P$ must commute with $T$, implies that
$P[\q\pm]$ must
be a sum of terms each with $p/2$ $\q +$'s and $p/2$ $\q -$'s.
(This is why we are limited to even $p$.)

Let $\omega$ denote the automorphism of $\uq$:
\eqn\IIxix{
\omega(T) = T,~~~~~
\omega (\q\pm ) = \qb\pm , ~~~~~\omega(\qb\pm ) = \q\pm , ~~~~~
\omega (q) = q^{-1} . }
Then given a solution $P$ to the above conditions, $\pb = \omega (P)$
is  a solution.

In the next section we will derive a field theory construction which
provides a powerful way of constructing solutions to the conditions
(i)-(iv).  This field theory construction deals only with elements that
may be expressed using multiple adjoint actions.
We emphasize however that the above algebraic
characterization
is overdetermined and if a solution exists one would expect it to
completely determine $P,\pb$.  Within the present algebraic
framework, the
existence of solutions to each of the conditions (i)-(iv) is only
possible due to the fact that $q^{2p} = 1$.
We have verified this by explicit computations for
$p=2,4,6$.   For the remainder of
this section, we describe some general properties of solutions and
outline the computations involved in determining them solely
algebraically.  As we will explain, the
conditions (iii) and (iv) are not completely independent.
We will present results for $p=2,4$ in this
section; results for $p=6,8$ will be presented in section 4.

It is known from the work of De Concini and Kac that the center of a
quantum group is enlarged when $q$ is  a root of unity\ref\kac{C. De
Concini
and V. G. Kac, Progress in Math. 92 (1990) 471. }.
The paper \kac\ was mainly concerned with the finite case, however
some
of its results are valid in the affine case as well.
The results in \kac\ are limited to a subset of the center which does
not contain elements of the type $P,\pb$.  Firstly, the paper \kac\
deals with an algebra $\uq'$ which is slightly different from
$\uq$.  In $\uq'$, $q^{\pm T} \equiv K_\pm $ are taken as basic
generators
rather than $T$ itself. In $\uq'$, as shown in \kac,  the elements
$K_\pm^p , Q^p_\pm , \Qb^p_\pm $ and their generalizations for other
positive roots are contained in the center.  Thus one sees that
$\uq$ and $\uq'$ are distinguished by $\uq$ having a $U(1)$
subalgebra,
and $\uq'$ only a $Z_p$ subalgebra.
Strictly speaking, $Q^p_\pm , \Qb^p_\pm$ are not central in $\uq$,
since
they do not commute with $T$.
In a physical realization of the symmetry $\uq '$,
since $Q^p_\pm , \Qb^p_\pm $ are central, they must either be
zero or proportional to the identity in specific representations.
In a theory with $U(1)$ symmetry, since $Q^p_\pm , \Qb^p_\pm$ are
not $U(1)$ invariant, they cannot be proportional to the
identity.  We will explain below that it
is algebraically consistent to have
\eqn\IIxx{
Q^p_\pm = \Qb^p_\pm = 0.}
Furthermore we will describe how \IIxx\ is proven in a
quantum field theory realization.
We will explain in section 6 how the other possibility
corresponding to $\uq'$ may be realized physically.
Thus the primary distinction between
our central elements  and the ones in \kac,
is that $P,\pb$ belong in the algebra $\uq$ and are relevent for
theories with $U(1)$ symmetry.
$P,\pb$ are also central in $\uq'$, and are to be associated
with imaginary roots.

Some remarks on the comultiplication condition (iv) are in order.
More generally, define an element $u \in \uq$ to be {\it primitive}
if it
has the following comultiplication
\eqn\IIxxi{
\Delta (u) = u\ot 1 + \Theta^{(u)} \ot u , }
for $\Th^{(u)} \in \uq$.  Primitive elements are important for the
following reasons.  Consider the Serre relation \IIii, and let
$u$ be the LHS of equation \IIii.  In order to consistently
impose $u=0$, one must have $\De (u) = 0$.  The latter is ensured by
the fact that $u$ is a primitive element with $\Th^{(u)} = q^{\pm
2T}$.
In fact, the precise form of the
LHS of the Serre relation is uniquely fixed by requiring
it to be primitive.
Primitive elements of $\uq$ are relatively rare.  For example, for
a general element $a$  of $\uq^+$ such as $a= Q_+^n$, one has
cross-terms
in the comultiplication of the form $\De (a) \propto .. +
Q_+^{n_1} \ot Q_+^{n_2}
+...$ with $n_1 + n_2 =n$ which cannot be re-expressed in terms of
$a$.

When $q$ is a root of unity, the number of primitive elements
increases.
It is not difficult to show for example that $Q^p_\pm$ are primitive
with $\Th = q^{\pm p T}$.  Again this primitive nature allows one to
consistently impose \IIxx.

It is very
important for us to note that primitive elements can be seen to be
automatically central under
certain conditions.
Suppose that $u$ is primitive, with $\Th = 1$, and furthermore
that $\ro (u) \propto 1$, where the constant of proportionality
may be zero, i.e. we suppose that $u$ is central in the particular
representation $\ro$.   Since all the higher dimensional
representations of
$\uq$ can be constructed on the space $V^{\ot N}$
by iterating the comultiplication\foot{See the discussion in section
6.},
the primitivity of $u$ ensures that
$u$ is central in all of these higher dimensional representations.
This implies that $u$ is central in the algebra.

Consider now the centrality condition itself, apart from the
primitivity
condition.
It is sufficient to check centrality with respect to a subset of
the generators of $\uq$.
Upon imposing $P=P^\dag$,
one has
\eqn\conse{
\a Q_+ (P) = \a \Qb_- (P) = 0 ~~\Rightarrow ~~
\a Q_- (P) = \a \Qb_+ (P) = 0 . }
Thus for Hermitian $P$, one only has to impose that it is cental
with respect to the finite subalgebra $\uqo$ generated by
$Q_+, \Qb_- , T$.
The only relations available for imposing $\a Q_+ (P) = 0$ are the
Serre relations.  One can argue that
\eqn\argue{
\a \Qb_- (P) = 0 ~~\Rightarrow~~
\a Q_+ (P) = 0, }
and vice-versa.
Recall that $\a \uqo$ is a $\uqo$ homomorphism.  This means that
the adjoint action of $\uqo$ on $P$ must define a representation
$\pi$
of $\uqo$.  Since $\a \Qb_- (P) = \a T (P) = 0$, and $Q_+^p = 0$,
$\pi$ is finite dimensional and   must be
the trivial one-dimensional representation.  Thus $\a Q_+ (P) = 0$.
To summarize, it is sufficient to impose
$\a \Qb_- (P) = 0$, the latter being easily computable from
\IIviii\ and \IIix.

For the cases which we have analyzed in detail, $P$ can be completely
determined by {\it either} imposing the primitivity or
$\a \Qb_- (P) =0$.  Again this illustrates how primitivity and
centrality are related.

We now outline the details of the
computations involved in determining $P$ from the
above algebraic structures.  One can begin by considering the most
general expression for $P$, which is a sum of
$\left(\matrix{p\cr p/2 \cr } \right)$ terms with arbitrary
coefficients,
each involving
$p/2$ $Q_+$'s and $p/2$ $Q_-$'s.
 For $p\geq 6$ it is important to realize that these
terms are not all linearly independent due to the Serre relations.
The problem of finding a linearly independent basis has
been solved in \ref\lustig{G. Lusztig, Geom. Ded. (1990) p 89.}.
The Hermiticity constraint is easily imposed and reduces the number
of independent coefficients.  A consequence of \IIxv\ is that
$S(P) = - P$, which is also straightforward to impose and further
reduces the number of undetermined coefficients.  The combined
constraints from Hermiticity and the antipode  are equivalent
to the constraints from CPT invariance.

{}From \IIiii\ and \IIiiib\ one can compute $\De (P)$.  Setting
to zero all  of the terms not of the form \IIxv\ yields an
overdetermined
system of linear equations for the coefficients.  For $p\geq 6$,
in doing this computation one must bear in mind that some of the
unwanted terms in $\De (P)$ may be automatically zero due to the
Serre relations, and this generally leads to weaker conditions on
the coefficients.  One can systematically take this into account by
re-expressing all terms in $\De (P)$ in terms of linearly
independent basis elements.
For $p=2,4,6$, due to the fact that $q^{2p} = 1$, one obtains
a unique solution up to an overall arbitrary constant (which we call
$c_p$ below).  For $p=2,4$
they are the following:
\eqn\IIxxiii{\eqalign{
P & ~~ =  ~~c_2 \(  \q+\q- + \q-\qp \)  =
c_2 \a \q+ (\q-) ~~~~~~~~~~~~~~~~~(p=2) \cr
P & ~~= ~~
c_4 \( (\q+ \q- )^2  - \q+ Q^2_- \q+ - Q_+^2 Q_-^2 - \q- Q^2_+ \q-
+ (\q- \q+ )^2  - Q^2_- Q^2_+  \) \cr
&=
\frac{c_4}{2} \>
\( \a (Q_+ \q- \q+ ) ( \q- ) - \a (Q^2_+  \q- )( \q- ) - \a (\q-
Q_+^2)
(\q- ) \),  ~~~ (p=4) \cr }}
For $p=4$ the primitivity of $P$ entails the cancelation of $70$
unwanted
cross-terms!

The centrality condition can be imposed as follows.
Imagine again starting
with a completely general expression for $P$ as a sum of linearly
independent terms with arbitary coefficients.
{}From \IIviii\ and \IIix\ one can easily
compute $\a \Qb_- (P)$, and setting the result to zero leads to
linear algebraic equations for the coefficients of the independent
terms in $P$.  Again, for $p\geq 6$, one must first re-express the
result of the computation of $\a \Qb_- (P)$ in a linearly independent
basis before setting it to zero, otherwise the resulting equations
are too strong.
Let us illustrate the centrality properties of $P$ for $p=2,4$.
The case of $p=2$ is in a certain sense degenerate, due to the fact
that
one cannot prove that $Q^2_\pm$ commute with $\q\pm$ from the Serre
relations.  However since $Q^2_\pm$ are primitive, it is consistent
to
set them to zero.  The Serre relations degenerate in this case since
they are now automatically satisfied.  One has that $[\q\pm , P] = 0$
due to \IIxx, and one can verify easily that $\a \Qb_\pm (P) = 0$,
using
the fact that $q^2= -1$.

The first non-degenerate case is $p=4$.  One can easily check that
$\ro (P) \propto  e^\th $ for the expressions in \IIxxiii.  Thus the
above arguments indicate that since $P$ is primitive, it must be
central.
Indeed, \IIxxiii\ is the unique solution to $\a \Qb_- (P) = 0$,
as verified by explicit computation.
The simplest way to show explicitly that $\a Q_+ (P) = 0$ is to use
the expression for $P$ in terms of $\a$'s.  Let $S_\pm$ denote the
LHS
of the first Serre relation in \IIii.  From the basic properties
\IIvii\ of the adjoint action, one has
\eqn\IIxxiv{
\a A \>  \a S_\pm \> \a B \, (C) = 0 ~~~~~~\forall \> A,B,C \in \uq .
}
Using this, one finds
\eqn\IIxxv{\eqalign{
\a Q_+ (P) &~\propto ~ \a \( Q^2_+ Q_- \q + - Q^3_+ \q - - \q+ \q-
Q^2_+ \)
\> \( Q_- \) \cr
&= -  ~ \a \q- \a^3 \q+ \( \q- \) = 0, \cr }}
where we have used \IIixb.

As we alluded to above, the main difficulty encountered for $p\geq 6$
is
that due to the Serre relations, there are linear relations among
elements of $\uq^+$, and one is forced to work with a linearly
independent
basis.   In dealing with this, we did not utilize the basis described
in \lustig, but rather adopted a simpler scheme which will be
described
when we present the solutions in section 4.
For $p\geq 8$, the above computations are intractible, and we rely
instead
on the field theory construction of the next section to generate
solutions.
As we will argue, the field theory formulation provides an efficient
means
of generating primitive elements. Furthermore, the field theory
construction
apparently automatically incorporates the linear dependencies of
elements
of $\uq^+$ due to the Serre relations.

Many physical systems which are known to have the $\uq$ algebra as
a symmetry (e.g. SG theory) are integrable, and thus have an infinite
number of commuting integrals of motion $I_n , \bar{I}_n$ of integer
Lorentz spin
$\pm n$.  If these higher integrals of motion are also contained in
$\uq$, then they must also satisfy the above conditions with
\IIxiii\ replaced by
\eqn\higher{
[L, I_n ] = n I_n , ~~~~~~~[L, \bar{I}_n ] = -n \bar{I}_n . }
The question arises as to whether we expect to be able
to construct infinitely many commuting integrals of motion
purely from the $\uq$ algebra.
In view of the fact that ordinary $N=2$ supersymmetry algebra
is a special case of $\uq$ one would expect that the answer
to the above question be no, since in the ordinary $N=2$ theory
the generic theory is {\it not} integrable, and so
we are not able to construct infinitely many integrals
of motion.
We expect the same is true for higher $p$, though the above
argument is not conclusive in these cases since, as
explained above, the $p=2$ point
is degenerate\foot{In the case of $p=4$ we searched unsuccessfully
for a spin 3 integral of motion as a sum of terms with 6 $Q_+$'s and
6 $Q_-$'s using the field theory construction of the next section.}.

However if the
exact form of $q$ is different from \IIxvi, the
conserved charge that we construct satisfies
all the requisite properties except that its spin, though integral,
is
different from that of $P$ and cannot be identified with it.
In particular consider instead of \IIxvi\
\eqn\pprime{
q^2 = \exp \( -2\pi i \frac{p'}{p} \) , ~~~~~~\Rightarrow s= p'/p }
where $p',p$ are relatively prime integers.
Since all one needs in the above construction is that $q^p = -1$,
then
$P$ is still a solution to (i),(iii),(iv), except that now it has
Lorentz spin $p'$ (which
is an odd integer), and can  be associated with the higher spin
integral of motion $I_{p'}$, in addition to the usual energy and momentum.
This shows that for example in the
case of the
SG model, where $q$ is a fixed function of coupling, one can
construct
integrals of motion of any odd integer spin, but one must vary the
coupling in order to do so.  See section 5.
The existence of this
additional higher integral of motion in a dense
subset of the coupling constant space strongly suggests that the
theory in integrable, but does not constitute a proof that this
is indeed the case.
Isolated theories  that are  $\uq$ invariant
with $q$ given in \IIxvi\ are not necessarily integrable.
Henceforth, unless
otherwise
stated, we are dealing with $p'=1$.

\newsec{Field Theory Formulation}

The $\uq$ symmetry of a 2d quantum field theory implies the existence
of
conserved currents $J^\mu (x)$, satisfying $\d_\mu J^\mu = 0$, for
each basic generator $Q$ of $\uq$, such that
$Q= \inv{2\pi i} \int_{-\infty}^\infty  dx \> J^t (x) $ is a
conserved
charge.  The conserved charge $T$ corresponds to a $U(1)$ symmetry of
the theory, and is generated by a local Lorentz spin 1 current.
Let  $J_\pm^\mu (x), \bar{J}_\pm^\mu (x); ~~\mu = z,\zb$ denote the
conserved
currents for the charges $\q\pm, \qb\pm$ respectively. The specific
construction of these currents is of course model-dependent. However
these currents are characterized by some fundamental
model-independent
features from which the Hopf algebra properties of $\uq$ follow.
We first review these facts, following \rbl.
For simplicity, we will focus on the currents $J^\mu_\pm$, however
all of the results in this section are valid with
$J_\pm \to \bar{J}_\pm , \q\pm \to \qb\pm , q\to q^{-1}, s\to -s $.

Let $\Psi (x)$ be a general field of the quantum field theory, and
let the braiding of the currents with these fields take the following
form
\eqn\IVi{
J_\pm^\mu (x,t) \> \Psi (y,t ) ~=~ q^{\pm T_\Psi} \>
\Psi(y,t) \> J^\mu_\pm (x,t ) , ~~~~~~x<y , }
where
$T_\Psi $ is the $U(1)$ charge of $\Psi$.
These braiding relations define a time-ordering prescription
$\hat{T}$:
\eqn\timeorder{\eqalign{
\hat{T} \( J_\pm^\mu (x,t+\ep ) \> \Psi (y) \) ~&=~
J_\pm^\mu (x, t+ \ep ) \> \Psi (y) ~~~~~~~~~~~~\ep > 0 \cr
&= q^{\pm T_\Psi } \> \Psi (y) \> J_\pm^\mu (x,t + \ep ) ~~~~~~\ep <
0 .
\cr }}
These braiding relations arise
due to the fact that the currents $J^\mu_\pm$ are generally
non-local.
Define the adjoint action of the charges $\q\pm$ on the field as
follows:
\eqn\IVii{
\a \q\pm \( \Psi (y) \) = \inv{2\pi i} \hat{T} \(
\oint_{\CC(y)} \> dx^\nu \varepsilon_{\nu\mu} \> J_\pm^\mu (x) \>
\Psi (y) \) }
where $\CC(y)$ is a contour beginning and ending at $x=-\infty$ and
surrounding the point $y$.  The precise shape of the contour is
irrelevant due to the conservation of the currents.  Break up
the contour $\CC(y)$ into two pieces $\CC_1 (y) + \CC_2 (y)$ where
$\CC_1$ is above $y$, in the time-ordered sense, and extends from
$-\infty$ to $+\infty$, whereas $\CC_2$ is below $y$ and goes
from $+\infty$ to $-\infty$.
The contour $\CC_1$ contributes $\q\pm \Psi (y)$ to \IVii, whereas
$\CC_2$ contributes $ - q^{\pm T_\Psi } \Psi (y) \q\pm $.  Thus
$\a \q\pm$ is a braided commutator:
\eqn\IViii{
\a \q\pm \( \Psi (y) \) = \q\pm \> \Psi (y) - q^{\pm  T_\Psi  } \>
\Psi (y)
\> \q\pm . }

If one takes $\Psi$ to be a current itself, and integrates over $y$
in
\IViii, then one sees that the adjoint action we have defined in the
quantum field theory is equivalent to the adjoint action \IIviii\ in
the algebra $\uq$.  The comultiplication \IIiii\ is also a
straightforward consequence of the braiding properties of the
currents.
One has
\eqn\IViv{
\a \q\pm \( \Psi (x) \Psi '(y) \)
= \a \q\pm \( \Psi (x) \) \> \Psi' (y)
+ q^{\pm  T_\Psi  } \>  \Psi (x) \>  \a \q\pm \( \Psi'(y) \) . }
This is shown by defining the LHS as in \IVii\ with the contour
$\CC$ surrounding both $x$ and $y$, and decomposing $\CC$ into two
separate contours, one surrounding $x$ the other $y$, and again
taking into account the time-ordering.  The equation \IViv\
translates directly into \IIiii\ if one interprets the two spaces
in the tensor product as the space of fields at the points $x$ and
$y$ respectively.  From the usual correspondence between fields
and the Hilbert space, one sees that the comultiplication in
\IIiii\ is valid on the multiparticle Hilbert space as well.  The
other
Hopf algebra properties also have a clear field theoretic
meaning\double\gs.
The antipode $S$ corresponds to a Euclidean rotation by an angle
$\pi$ in the $x-t$ plane, and the counit $\ep$ is the 1-dimensional
vacuum representation of $\uq$.

Let $L$ be the generator of Lorentz boosts, and let the spin
$s$ of the charges be defined by $[L,\q\pm ] = s\> \q\pm $.
For the currents $J_\pm$, the braiding relations \IVi\ read
\eqn\IVv{
J^\mu_\ep (x) \> J^\nu_{\ep'} (y) = q^{\ep\ep' 2} ~
J^\nu_{\ep'} (y) \> J^\mu_{\ep} (x)  ~~~~~~~x<y  , }
($\ep ,\ep ' = \pm $).
If $q^2 = \exp (-2\pi i \al )$, then from the braiding relations
\IVv\ with $\ep = \ep'$, one deduces that $s=\al ~ {\rm Mod }\Zmath$.

We remark that it follows from the above formulation that for local
conserved charges and  currents with trivial braiding, $\De$
is the trivial one, and the adjoint action is the usual commutator.

In order to make contact with the previous section, one needs to
consider multiple adjoint actions of charges on fields. Consider
the expression
\eqn\IVvi{
\a \q{a_1} \a \q{a_2} \cdots \a \q{a_{n-1}}
\( J_{a_n}^\mu (y) \), }
where
$a_i \in \{ \pm \} $. In general one cannot shrink the contours in
this expression due to presence of multiple cuts.
However under certain special circumstances the contours in
\IVvi\ may be shrunk arbitrarily close to $y$, and the operator
product expansion used, to yield a new meaningful operator.
In this way one can obtain a local operator, such as the
energy-momentum
tensor.
These contour shrinkage conditions were studied
in \ref\rfl{G. Felder and A. LeClair,
Int. Journ. Mod. Phys. A7 Suppl. 1A (1992) 239.},
based on ideas introduced in
\ref\rfw{G. Felder and C. Wieczerkowski, Commun. Math.
Phys. 131 (1990) 125.}.  In \rfl, the contour shrinking conditions
were
derived in the special case where $a_1 = a_2 = ...= a_{n-1}$.  We now
generalize this result.

To simplify the discussion we introduce a vector space notation.  Let
$W \sim
\Cmath^2$
denote the 2-dimensional vector space corresponding to the $\pm$
indices of
the charges $\q\pm$.  For reasons that will become clear, we express
the
braiding relations \IVv\ as follows
\eqn\braid{
J_1 (x) \> J_2 (y) ~=~ \rh_{12} \> J_1 (y) J_2 (x) ~~~~~~~~~~x<y, }
where
$\rh_{12}$ is a $4\times 4$ matrix acting on $W\ot W$, and the
subscripts label individual $W$ components of $W\ot W$.
Specifically,
\eqn\er{
\rh_{12} = R_{12} \> \sigma_{12} , }
where
$R_{12} = {\rm diag} (q^2,q^{-2},q^{-2},q^2 )$, and $\sigma_{12}$ is
the operator that permutes two spaces:
\eqn\perm{
\sigma_{12} = \left( \matrix{1&0&0&0\cr0&0&1&0\cr0&1&0&0\cr0&0&0&1 \cr }
\right)
 . }
The expression \IVvi\ may then be viewed as a specific vector in the
$2^n$ dimensional vector space $W^{\ot n}$.

Consider now the expression
\eqn\IVvii{
\a \q{1} \a \q{2} \cdots \a \q{{n-1}} \( J^\mu_n (y) \)
= \inv{(2\pi i)^{n-1}}
 \hat{T}  \( \prod_{i=1}^{n-1} \oint_{\CC_i}
dx_i J^*_{i} (x_i) \) J^\mu_n (y) }
where
$J^{*\mu} \equiv \ep^\mu_\nu J^\nu$.
The above expression is viewed as a general vector $v\in W^{\ot n}$,
where the subscipts label one of the components $W$ of the
n-fold tensor product.  All the contours $\CC_i$ begin and end at
$x_\infty \equiv - \infty$, and surround $y$ as shown in figure
1.
\midinsert
\epsfxsize = 3in
\bigskip\bigskip\bigskip\bigskip
\vbox{\vskip -.1in\hbox{\centerline{\epsffile{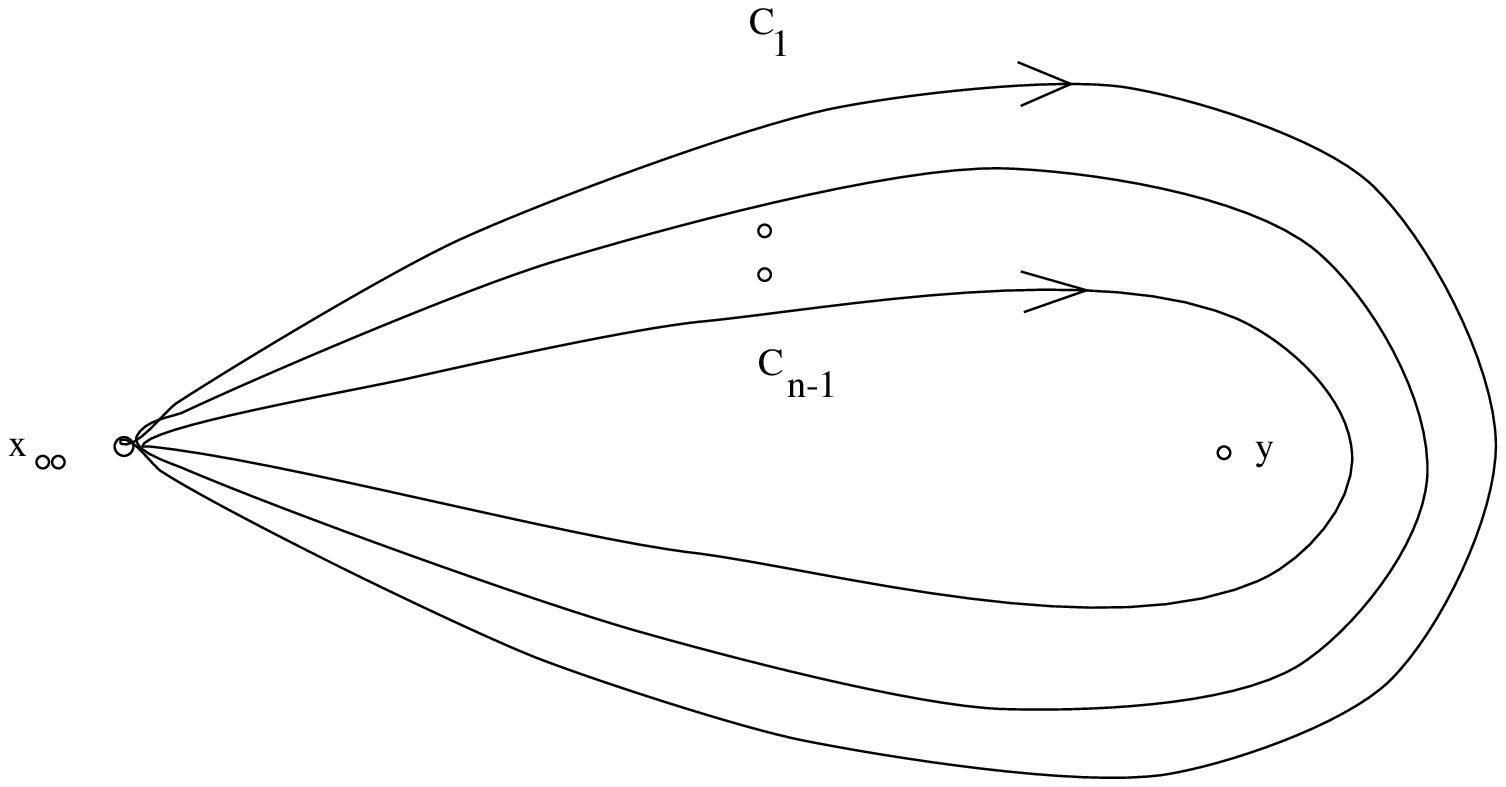}}}
\vskip .1in
{\leftskip .5in \rightskip .5in \noindent \ninerm \baselineskip=10pt
Figure 1.
The contours representing the expression \IVvii. The time and space
axes run
vertically and horizontally respectively.
\smallskip}} \bigskip
\endinsert

\def\inf{{\infty}}

Now consider deforming the point $x_\infty \to x_\infty + \ep$,
where $\ep$ is some infinitestimal vector.
In general, the point $x_\infty$ cannot be moved due to
the presence of multiple cuts.
To derive the conditions underwhich $x_\infty$ can be moved,
we compute  the derivative with respect to $\ep$ of the expression in
\IVvii.  When this is zero,  then the contours may be shrunk to a
region
arbitrarily close to $y$.  The derivative with respect to $\ep$
of the expression \IVvii\ yields a sum of $2(n-1)$ terms, which
arise from the two boundaries of each integral $\oint dx_i$.
Let $J(x_\infty^+)$ and $J(x_\infty^- )$ respectively
denote the currents at the upper and lower boundaries of each
contour integral.  As space-time points, $x_\infty^+ = x_\infty^-$,
however the two fields $J(x_\infty^+ )$ and $J(x_\infty^- )$ are
essentially different due to the non-locality of the currents.
For concreteness consider first the contributions to $\d_\ep$ coming
from $\oint dx_1$:
\eqn\IVviii{
\( J_1^* (x_\infty^+ ) - J_1^* (x_\infty^- ) \)
{}~
\[ \prod_{i=2}^{n-1} \oint d x_i \> J^*_i (x_i ) \] ~ J^*_n (y) . }
In order to understand this expression, it is useful to represent the
non-locality of the currents with a string which runs from
$-\infty$ to the location of the current and represents the cut
associated to the current; this string is analagous to the disorder
line
defining disorder fields in 2 dimensions, and is  manifest in the
example of the SG theory.  The fields $J_1^* (x_\infty^\pm )$ in
the expression \IVviii\ are represented graphically in figure 2.
\midinsert
\epsfxsize = 3in
\bigskip\bigskip\bigskip\bigskip
\vbox{\vskip -.1in\hbox{\centerline{\epsffile{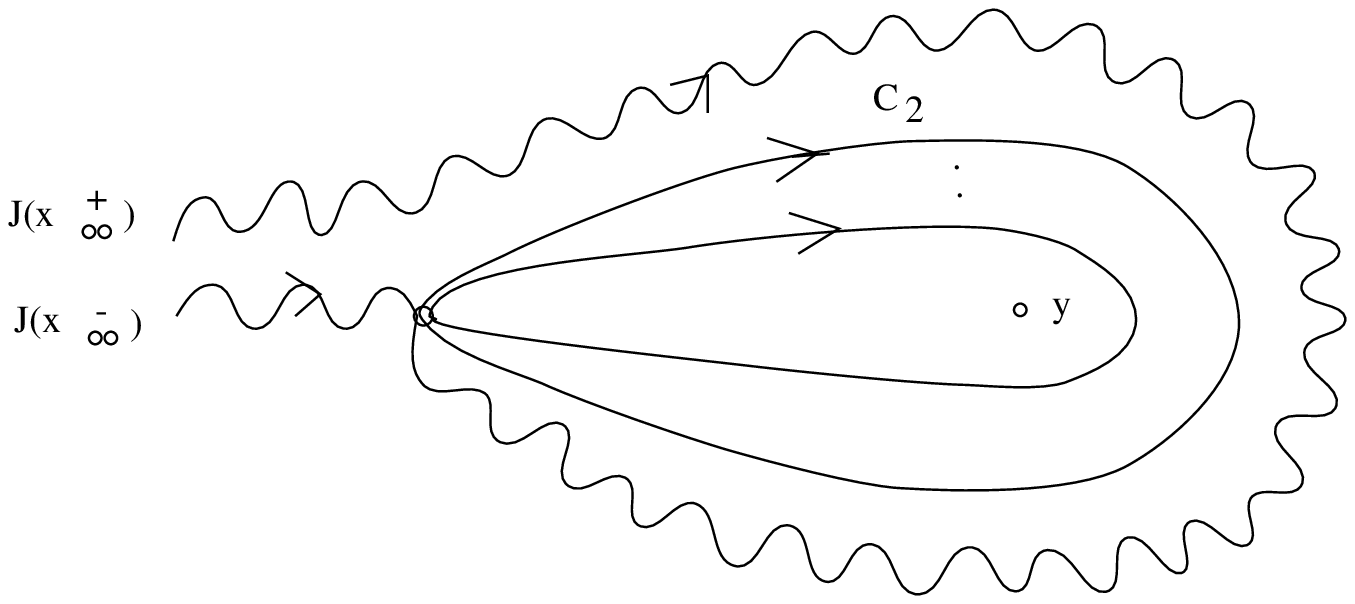}}}
\vskip .1in
{\leftskip .5in \rightskip .5in \noindent \ninerm \baselineskip=10pt
Figure 2.
Graphical representation of equation \IVviii, where wavy lines
denote the strings.
\smallskip}} \bigskip
\endinsert
One sees that the two terms in \IVviii\ are equivalent up to braiding
factors
that arise when $J_1^* (x_\infty^+ )$ completely encircles the
fields $J_i^* (x_i ), i=2,..,n$.
Thus eq. \IVviii\ can be rewritten as
\eqn\IVix{\eqalign{
& \( \rh_{12} \rh_{23} \cdots \rh_{n-2,n-1} \rh^2_{n-1,n}
\rh_{n-2,n-1} \cdots \rh_{23} \rh_{12} - 1 \) \cdot \cr
&~~~~~~~~~~~~~J_1^* (x_\inf )
\( \prod_{i=2}^{n-1} \oint dx_i J_i^* (x_i ) \) \> J_n^* (y) . \cr}}

All of the other terms in $\d_\ep$ of \IVvii\ are of the form of the
term in \IVix\ up to braiding factors.  Consider for example the 2
terms
arising from the boundaries of $\oint dx_2$:
\eqn\IVx{
\oint dx_1 J_1^* (x_1 ) \[ J_2 (x_\inf^+ ) - J_2 (x_\inf^- ) \]
\( \prod_{i=3}^{n-1} \oint dx_i J_i^* (x_i ) \) \> J_n^\mu (y) . }
As above, the current $J_2 (x_\inf^+ )$ can be encircled back around
$J_i , i= 3,..,n$ yielding the factor
$$ \rh_{23} \rh_{34} \cdots \rh_{n-2,n-1} \rh^2_{n-1,n} \rh_{n-2,n-1}
\cdots \rh_{34} \rh_{23} $$
so that the two terms in \IVx\ are of the same type. Then to obtain a
term
of the type in \IVix\ the current $J_1 (x_1)$ can be braided through
$J_2$
and the change of integration variables $x_1 \to x_2 $ performed.
One obtains for \IVx
\eqn\IVxi{\eqalign{
\rh_{12} & \(  \rh_{23} \rh_{34} \cdots \rh_{n-2,n-1} \rh^2_{n-1,n}
\rh_{n-2,n-1}
\cdots \rh_{34} \rh_{23}  -1 \) \cdot
\cr
& ~~~~~~~~~~~~~~~J_1^* (x_\infty )
\( \prod_{i=2}^{n-1} \oint dx_i J_i^* (x_i ) \) \> J_n^* (y) . \cr}}
Putting all of this together, one finds
\eqn\IVxii{
- \der{\ep} ~ \a \q{1} \a \q{2} \cdots \a \q{{n-1}} \( J^\mu_n (y) \)
=
M ~
J_1 (x_\infty^- )
\a \q{2} \cdots \a \q{{n-1}} \( J^\mu_n (y) \)
,}
where
$M$ is a matrix acting on $W^{\ot n} \to W^{\ot n}$:
\eqn\IVxiii{\eqalign{
M & \equiv \CP - B \tilde{\CP} \cr
\CP &= 1 + \rh_{12} + \rh_{12} \rh_{23} + \ldots + \rh_{12} \rh_{23}
\cdots \rh_{n-2,n-1} \cr
\tilde{\CP} &= 1 + \rh_{n-2,n-1} + \rh_{n-2,n-1} \rh_{n-3,n-2} +
\ldots + \rh_{n-2,n-1} \cdots \rh_{12} \cr
B &= \rh_{12} \rh_{23} \cdots \rh_{n-2,n-1} \rh^2_{n-1,n} . \cr}}

The conditions underwhich the contours may be shrunk are thereby
reduced
to the following problem in linear algebra.  Denote by
$v(a_1 ,a_2 , \ldots ,a_n ), a_i = \pm$ the obvious basis vectors of
$W^{\ot n} $.  For any vector $v$ define
the field $J_v (y)$
such that for the
basis vectors one has
\eqn\IVxv{
J_{v(a_1 .. a_n )} \equiv
\a \q{a_1} \cdots \a \q{a_{n-1}} \( J^\mu_{a_n}  (y) \). }
Similarly define $Q_v \in \uq$ by
\eqn\IVxvi{
Q_{v(a_1 ..., a_n ) } \equiv
\a \q{a_1} \cdots \a \q{a_{n-1}} \( Q_{a_n}  \). }
One has $Q_v = \inv{2\pi i} \int_{-\infty}^\infty dy J_v (y)$.  Given
a null vector $v^{(0)}$ of $M$, $M\> v^{(0)} = 0$, then
equation \IVxii\ shows that the contours may be shrunk in
the expression $J_{v^{(0)}}$.  The resulting current in
conserved, as is the charge $Q_{v^{(0)}}$.

The above contour shrinking condition is actually too strong,
due to the existence of the Serre relations which imply that some
of the terms on the RHS of \IVxii\ may automatically be zero.
We now describe the weaker form.  For $v\in W^{\ot n}$ define
$\hat{v}\in W^{\ot (n-1)}$ as the evident projection of
$v$ onto its last $n-1$ components $W_2 \ot \cdots \ot W_n$.
Let $Q_{\hat{v}} \in \uq$ be defined as in \IVxvi. For some
$\hat{v}$, $Q_{\hat{v}}$ will equal zero due to the Serre relations.
Define $W_0^{(n)} \subset W^{\ot n}$ as
\eqn\IVxvii{
W_0^{(n)} \equiv \{ v\in W^{\ot n} : Q_{\hat{v}} = 0 \} . }
We state our final result.  The contours can be shrunk in the
expression $J_{v^{(0)}}$ if
\eqn\IVxviii{
M \> v^{(0)} = 0 ~~~~~{\rm Mod } \, W_0 . }

The contour shrinking condition \IVxviii\ has a definite algebraic
significance.  Consider the expression
\eqn\IVxix{
\a Q_{v^{(0)}} \( \Psi (y) \> \Psi' (y' ) \)
= \inv{2\pi i} \int d x^\mu \ep_{\mu\nu} J^\nu_{v^{(0)}} (x) \>
\Psi (y) \Psi ' (y') , }
where as usual one integrates the current $J_{v^{(0)}} (x)$ along a
contour
surrounding $y$ and $y'$.  Since the multiple contours in the
definition of
$J_{v^{(0)}} (x)$ can be shrunk arbitarily close to $x$, one can
safely decompose the $x$-contour in \IVxix\ into two pieces, one
surrounding $y$ the other $y'$ to obtain
\eqn\IVxx{
\a Q_{v^{(0)}} \( \Psi (y) \> \Psi' (y') \)
=
\a Q_{v^{(0)}} \( \Psi (y) \) \> \Psi' (y') +
\a \Th \( \Psi(y) \) \a Q_{v^{(0)}} \( \Psi' (y') \) , }
for some $\Th \in \uq$.  If the contours could not be
shrunk in the definition of $J_{v^{(0)}}$, one would obtain
additional terms in \IVxx\ of the type $\a A \( \Psi (y) \)
\a B  \( \Psi '(y') \)$, where $A,B \in \uq$ are of lower degree
than $Q_{v^{(0)}}$.  Thus the contour shrinking condition ensures
$Q_{v^{(0)}}$ is a primitive element.

We remark that for local currents, where $R_{12} =1$, {\it any}
vector
$v\in W^{\ot n}$ satisfies \IVxviii.  Algebraically this corresponds
to the fact that in this case any element of the form
$[Q_{a_1} , [Q_{a_2} ,\cdots [ Q_{a_{n-1}}, Q_{a_n} ]\cdot\cdot]]$
is primitive.

\def\vo{{v^{(0)} }}

We first illustrate the above construction with some simple examples.
Consider first the special case where all of the charges $Q_{a_i} ,
i=1,..,n-1$ are the same, so that we consider solutions of \IVxviii\
of
the form $v^{(0)} = v(a,a,...,a, a_n ), a=+ ~{\rm or}~-$.  Note that
for
$v^{(0)}$ of this type, $\CP v^{(0)} = \tilde{\CP} \vo \propto \vo$,
where the constant of proportionality is a sum of phases.
Therefore, in this case the contour shrinking condition is reduced
to
\eqn\IVxxi{
(1-B) \vo = 0 .}
The simplest solutions to \IVxxi\ for generic $q$ are the vectors
$\vo = v(+,+,+,-)$ and $ v(-,-,-,+)$.  These solutions allow one to
prove
the Serre relations, as was done in \rfl. Namely,
the above arguments show that $J_{v(\pm\pm\pm \mp )}
= \a^3 \q\pm \( J_\mp \)$ are well defined operators in the quantum
field theory.  By using (model-dependent) scaling arguments, one
shows
that
$\a^3 \q\pm (J_\mp ) = 0$.
For generic $q$, there exist other solutions of \IVxviii\ for $n>4$.
However they do not contain anything new, since for these other
solutions
$\vo$, $Q_\vo = 0$ as a consequence of the Serre relations.  For
example
$\a^m \q\pm (\q\mp ) = 0$, $m\geq 3$.

When $q$ is a root of unity, there are new solutions to the
contour shrinking conditions.
As before let $q^{2p} = 1$, such that the
charges $\q\pm$ have Lorentz spin $1/p$.
The simplest example is
the following.
Consider
the vector $\vo = v(+,+,+,..,+,+) \in W^{\ot p}$ with
all $+$'s. One finds that $B \vo = q^{2p} \vo$.  Thus the contours
can be shrunk in the expression $\a^{p-1} \q+ \( J_+ (y) \)$ and
similarly
for the expression $\a^{p-1} \q - \( J_- (y) \)$.
This implies that we have  charged conserved integrals of
motion $Q_{\pm}^p$.  Since in the generic theory one does
not expect to have any {\it extra} integrals of motion other
than the energy and momentum, which are neutral, these
operators better vanish identically. In particular,
in section 5, within the context of the SG model, we will use scaling
arguments to show that
\eqn\IVxxii{
\a^{p-1} \q\pm \( J_\pm \) = 0 ~~~\Rightarrow  Q^p_\pm = 0 . }

We now apply the above construction to the momentum operators $P$.
Since the adjoint action involves mixed $Q_+ , Q_-$'s, there is no
simplification of \IVxviii.  First consider the case of $p=4$.
Here $W_0^{(4)} = \emptyset$.
One finds a unique solution to \IVxviii\ (up to an overall arbitrary
constant):
\eqn\IVxxiii{
\vo = v(+-+-) - v(++--) - v(-++-) ~~~~~~~(p=4) . }
The conserved current $J_\vo$ has Lorentz spin 2.  In the specific
case
of the SG theory, $J_\vo$ may be identified with the component of
the energy-momentum
tensor $\CT (y)$:
\eqn\IVxxiv{
\a \( \q + Q_- \q +  - Q^2_+ Q_- - Q_- Q^2_+ \) \( J^\mu_- (y) \)
\propto \CT^\mu_z  (y) .}
Integration of this equation yields \IIxxiii.

\def\vot{\tilde{v}^{(0)} }

Solutions to the contour shrinking conditions for $p=6,8$ will
be presented in the next section.  We finish this section with
some remarks.  In general, the solutions to \IVxviii\ are  not
unique.  For the cases we have examined in detail,  this multiplicity
of solutions  corresponds to the same element $P$ as a consequence
of the Serre relations.  Given an element of $\uq^+$ such as
$P$, generally there is no unique expression for it in terms of
$\a$'s due to the identity \IIxxiv.  We assert without proof the
following:
Given a solution $\vo$ to \IVxviii\ and its associated
$Q_{\vo}$,
then any other $\vot$ such that due to the Serre relations
$Q_{\vot} = Q_{\vo}$ is also a solution.
For example, at $p=6$ we obtained a 4-parameter family of
solutions to \IVxviii, and all are equivalent due to the
Serre relations (see section 4).
Thus the field theory approach to the construction of $P$
automatically resolves some of the complications which arise
due to the linear dependencies of basis elements in the
algebraic approach.

\newsec{Explicit Solutions for $p=6,8$}

As stated previously, for $p\geq 6$, one must work in a linearly
independent basis of $\uq$ in order to construct $P,\pb$ from
the algebraic construction of section 2.
The scheme we adopted for finding a linearly independent basis
is the following.  Consider a general expression in $\uq^+$.  The
Serre relations \IIii\ can be used to reexpress any terms involving
3 consecutive $Q_+$'s or $Q_-$'s as a sum of terms which do not.
In this way, we constucted a basis in $\uq^+$ with the $Q_-$'s to the
left of the expression as much as possible, by repeatedly using
the identities:
\eqn\Vi{\eqalign{
A\> Q^3_+ Q_- \> B &= A \( (1+q^2 + q^{-2} )
Q^2_+ \q - \q + - (1 + q^2 + q^{-2} ) Q_+ Q_- Q^2_+ + Q_- Q^3_+ \) B
\cr
A\> Q_+ Q^3_- \> B &= A \(
(1+q^2 + q^{-2} ) Q_- Q_+ Q^2_- - (1 + q^2 + q^{-2} ) Q_-^2 Q_+ Q_-
+ Q^3_- Q_+ \) B  . \cr }}
The same scheme can be used to remove linear dependencies amoung
elements
of $\uq$ that are expressed in terms of $\a$'s.
One has $ \a L \, (C) = \a R \, (C)$
where $L,R$ are the
left and right hand sides of eq. \Vi.

Though the above scheme has served for our purposes, it has the
drawback
of  not
completely removing linear dependencies.  For example, one can
re-express
$Q_+^3 Q_-^3$ by either moving the $Q_+$'s to the right using the
first
relation in \Vi, or alternatively move the $Q_-$'s to the left using
the second relation.  The resulting expressions are not identical.
Setting them equal, one obtains the identity:
\eqn\Vii{\eqalign{
\q+ &Q^2_- \q+ \q- \q+ - \q- \q+ Q^2_- Q^2_+ - \q- \q+ \q- Q^2_+ \q -
- \qp \qm\qp Q^2_- \q + \cr
& +Q^2_+ Q^2_- \q + \q -  - \qp\qm Q^2_+ Q^2_-
+ Q_- Q^2_+ \q- \q+ \q- + Q^2_- Q^2_+ Q_- Q_+ = 0 . \cr }}

We now outline the analysis of the contour shrinking conditions
required for the construction of $P$ at $p=6$.  The action of
the matrix $M$ in \IVxviii\ on basis vectors $v(a_1,..,a_n)$
leaves the index $a_n$ unchanged, therefore it is sufficient to
study solutions  on the vector space spanned by basis vectors
with $a_n = -$.  The most general vector $\vo$  of this type
with $n=6$ consists of a sum of 10 terms. The ${\rm Mod}\> W_0$
condition may be imposed as follows. Due to the Serre relations,
$$Q_{\hat{v} (--+++-)} = 0, ~~~~
Q_{\hat{v} (-+++--)} = 2 \> Q_{\hat{v} (-++-+-)} - 2 \> Q_{\hat{v}
(-+-++-)}.$$
Thus one should make the replacements
$v(--+++-) = 0$ and $v(-+++--) = 2 v(-++-+-) -2 v(-+-++-) $
after computing $M\> \vo$ and before setting the result to zero.
In this way one obtains 8 linear equations for the 10 unknown
coefficients in the vector $\vo$.  After a long but
straightforward computation, one finds a 4-parameter family
of solutions:
\eqn\Viii{\eqalign{
v^{(0)} &= 2a\>  v(+-+-+-) + 2b \> v(+++---) + 2c \>  v(-+++--) \cr
& ~~~~~+ 2d \> v(--+++-) + (a-4b) \> v(++-+--)
\cr & ~~~~~ +(2a-4b-4c) \> v(-++-+-) +(4b-2a ) \>  v(+-++--)
\cr & ~~~~~ + (4b+4c -4a) \> v(-+-++-) - a\> v(++--+-) . \cr}}
Remarkably, for arbitrary  constants $a,b,c,d$, $Q_\vo$
represents a
unique element of $\uq$ up to an overall constant.
This is because the terms in $Q_{v^{(0)}}$ proportional to $b,c$, or
$d$ are
all zero due to the Serre relations.  This example illustrates
how the contour shrinking conditions automatically incorporate
Serre's consequences.  Indeed, one may find solutions to $M\> \vo =0$
ignoring the ${\rm Mod} \> W_0$ condition altogether.  The
result of this computation is a two parameter family of solutions,
which are again equivalent up to Serre relations, and are the same as
\Viii.  Setting $b=c=d=0$,
and identifying $Q_\vo$ with $P$, one obtains the basic expression
\eqn\Viv{\eqalign{
P ~~ = ~~ \frac{c_6}{6} \>
&\a ( 2 \q + \q - \q + \q- \q+ + Q^2_+ \q - \q + \q-
+ 2 Q_- Q^2_+ Q_- Q_+ \cr
& ~~~~~-2 Q_+ Q_- Q^2_+ Q_- - 4 Q_- Q_+ Q_- Q_+^2
-Q^2_+ Q^2_- Q_+ ) ~\( Q_- \) . \cr }}

Writing out the $\a$'s in \Viv\ and reexpressing the result in a
linearly
independent basis using \Vi\Vii, one obtains
\eqn\Vv{\eqalign{
P ~~= ~~& \frac{c_6}{2} \>   (
2(\q+ \q-)^3   + 2 (\q- \q+)^3
+Q^2_+ \q- \q+ Q^2_- - 2 \q- Q^2_+ \q- \q+ \q- \cr
&~~~~~
+ 2\q+ \q- Q^2_+ Q^2_- - 5 Q^2_+ Q^2_- \q+ \q-
-9 Q^2_- \q+ \q- Q^2_+ - 7 \q+ Q^2_- \q+ \q- \q+ \cr
&~~~~~~~~~~
+ 7 \q- \q+ Q^2_- Q^2_+ + 5 \q+ \q- \q+ Q^2_- \q+
+ 5 Q^3_- Q^3_+ ) . \cr }}
It is simple to check that $\ro (P) \propto  e^\th  $, as only
the first two terms in \Vv\ contribute.
We have checked explicitly that $P$ is hermitian, primitive,
and central, in the way described in section 2.  One must use
the Serre relations to verify each of these properties.

\def\sq{\sqrt{2}}

For $p=8$ we found the following solution to the contour
shrinking condition:
\eqn\peight{
P ~~\propto ~~ \a Y \> (\q- )  ~~~~~~~~~~~~~~~(p=8) , }
where
\eqn\peightb{\eqalign{
Y &= 2 \q- Q^2_+ Q^2_- Q^2_+ - (6+4\sqrt{2} ) \q- Q^2_+ (\q- \q+)^2
+ (2+ 2\sqrt{2} ) \q- Q^2_+ \q- Q^2_+ \q-  \cr
&~~~~~-2 \q+ \q- \q+ Q^2_- Q^2_+
+(6+4\sq ) \q+ \q- \q+ \q- \q+ \q- \q+
\cr &~~~~~
+(4+4\sq ) \q+ \q- Q^2_+ Q^2_- \q+
- 2\sq Q^2_+ Q^2_- \q+ \q- \q+
+ 2 Q^2_+ Q^2_- Q^2_+ \q-
\cr &~~~~~
- (10 + 4\sq ) Q^2_+ \q- \q+ Q^2_- \q+
- 4 Q^2_+ \q- Q^2_+ Q^2_-
+ (7\sq - 2 ) Q^3_+ Q^3_- \q+
\cr &~~~~~
+(\sq - 2 ) Q^3_+ Q^2_- \q+ \q-
+ (6+\sq ) Q^3_+ \q- \q+ Q^2_-
- 3 \sq Q^4_+ Q^3_- \cr
&~~~~~- ( 2 + 2\sq ) \q+ \q- \q+ \q- Q^2_+ \q-
\cr}}
Due to the voluminousness of the computations involved, we  have
not checked that the above $P$ satisfies the requisite algebraic
properties described in section 2.  However we did verify that
$P$ is proportional to the identity  in the 2-dimensional
representation.

\newsec{Generalized Supersymmetries in the Sine-Gordon Theory}

In this section we describe how the above structures are
realized in a specific model, namely the SG theory.
The following discussion generalizes easily to the fractional
supersymmetric SG theories\rbl.
The SG theory is defined by the action
\eqn\ei{
S= \inv{4\pi} \int ~ d^2 z
\( \d_z \Phi \d_\zb \Phi + 4 \la : \cos (\be \Phi ):  \) . }
With the above normalization of the kinetic term\foot{
With our normalization of the kinetic term, the free fermion
point occurs at $\be = 1$. }, one has
$$< \Phi (z,\zb ) \Phi (0) > = - \log (z\zb ) .$$
The coupling constant has scaling dimension $2-\be^2$.
For $0\leq \be^2 \leq 2$, the theory may be viewed as a
relevant perturbation of the $c=1$ conformal field theory
of a single real scalar field, where the conformal field theory
is recovered in the ultraviolet (u.v.) limit $\la \to 0$.

The SG theory is known to possess a $\uq$ symmetry,
with
$$ q= \exp \( -2\pi i /\be^2 \)$$
which we now review\rbl.  The generator $T$ is the usual topological
charge
\eqn\eii{
T = \frac{\be}{2\pi} \int_{-\infty}^{+\infty} dx \> \d_x \Phi . }
The conserved currents $J_\pm^\mu , \bar{J}_\pm^\mu$ for the
charges $\q\pm , \qb\pm$ are non-local and take the following form
\eqn\eiii{\eqalign{
J_{\pm , z} (z,\zb) &= \exp \( \pm \frac{2i}{\be} \phi  \) , ~~~~
J_{\pm,\zb} (z,\zb) = \la \gamma \exp \( \pm i (\frac{2}{\be} - \be )
\phi
\mp i\be \phib \) \cr
\bar{J}_{\pm ,\zb} (z, \zb) &= \exp \( \mp \frac{2i}\be  \phib  \) ,
{}~~~~
\bar{J}_{\pm ,z} (z,\zb)
= \la \gamma \exp \( \mp i (\frac2\be  - \be )\phib \pm i \be
\phi \) , \cr}}
where $\gamma\equiv \be^2 /(2-\be^2 )$, and $\phi , \phib$ are the
quasichiral\foot{In this context
`chiral' refers to the distinction between left and right
movers
in the conformal limit.} components of the scalar field $\Phi = \phi
+ \phib$,
\eqn\eiv{\eqalign{
\phi (x,t) &= \inv{2} \( \Phi (x,t) + \int_{-\infty}^x dy \d_t \Phi
(y,t) \)
\cr
\phib (x,t) &= \inv{2} \( \Phi (x,t) - \int_{-\infty}^x dy \d_t \Phi
(y,t) \).
\cr}}
The conservation of these currents is established
to all orders in $\la$ using conformal perturbation
theory of the kind developed in \ref\zamo{A. B. Zamolodchikov,
Int. Journ. Mod. Phys. A4 (1989) 4235;  Adv. Studies in Pure
Math, vol. 19 (1989) 641.}, and is
primarily a consequence of the relations
\eqn\eres{\eqalign{
{\rm res}_{z=w} ~ ~ J_{+,z} (z) \> \exp ( -i \be \phi (w) )
&= \gamma \d_z \chi^* (z) \cr
{\rm res}_{z=w} ~ ~ J_{-,z} (z) \> \exp ( +i \be \phi (w) )
&= \gamma \d_z \chi (z) , \cr }}
where
$$\chi = \exp \( i (\be - 2/\be ) \phi \) , ~~~~~\chi^* = \exp \(
-i (\be - 2/ \be ) \phi \), $$
and similar relations for $\bar{J}_\pm$.
{}From the conformal scaling dimensions of the currents, one finds
that
the charges $\q\pm$ and $\qb\pm$ have Lorentz spin
$s=2/\be^2 - 1$ and $ 1- 2/ \be^2 $ respectively.

The last relation in \IIi\ is a consequence of the following
relations
\eqn\xtop{\eqalign{
\q+\qb- - q^{-2} \qb- \q+
&= \frac{\la}{2\pi i} \gamma^2 \int dx ~ \d_x X^*
 \cr
\qb+ \q-  - q^{2}  \q- \qb+
&=  - \frac{\la}{2\pi i} \gamma^2 \int dx ~ \d_x X ,
\cr}}
where $X$ is the local spinless field
\eqn\IIx{
X(z,\zb) = \chi \bar{\chi} = \exp \( i (\be - \frac2\be ) \Phi \) , }
and $X^*$ its complex conjugate.  We will use this result in the next
section.

The spectrum of the theory contains soliton doublets of
topological charge $T= \pm 1$.  Denote by $|\th ,\pm >$ the
1-soliton states of rapidity $\th$, where as usual
\eqn\eviii{
P =  m \, e^\th , ~~~~~\pb =  m\, e^{-\th} . }
The representation $\ro$ of $\uq$ on the one-soliton states is as
in \IIxxii.  The complete spectrum for all $\be$ and the S-matrices
are known\ref\rzz{A. B. Zamolodchikov and Al. B. Zamolodchikov,
Ann. Phys. 120 (1979) 253.}.

At the special values of the coupling
\eqn\eix{
\be = \sqrt{ \frac{2p}{p+1} } , }
one has
$q = - \exp ( -i\pi / p )$, and the charges $\q\pm , \qb\pm$ have
Lorentz spin $\pm 1/p$.  All of the general arguments of the previous
sections apply.  To complete the identification, we provide the
necessary scaling arguments.

Let us assume that all fields in the SG theory have a smooth u.v.
limit where they can be identified with fields in the $c=1$ conformal
field theory, as is normally done in conformal perturbation
theory\zamo.
Consider first the relation $Q_+^p = 0$.  The results of the previous
section show that $\a^{p-1} Q_+ (J_+ )$ is a well-defined field of
scaling dimension 2.  In the u.v. limit as $\la \to 0$, this field
must be in the chiral conformal family of the field $\exp
(\frac{2pi}{\be}
\phi )$.  All fields in this family have dimension greater than or
equal to $p(p+1)$, thus to zero-th order in $\la$,
$\a^{p-1} \q+ (J_+ ) = 0$.  From the form of the currents
$J_+$ \eiii, one infers that higher order contributions in $\la$
to $\a^{p-1} \q+ (J_+ ) $ must be in the conformal family of
the fields $\exp \( \frac{2pi}{\be} + i n \be (\phi + \phib ) \)$ for
any integer $n$.  Since the dimensions of these fields are greater
than
or equal to $p(p+1)$, and since $\la$ has positive scaling dimension,
one concludes that $\a^{p-1} \q+ (J_+ ) = 0$ to all orders in $\la$.
Similar arguments apply to the other relations in \IVxxii.

Next consider the relations involving $P$.  Let $\vo$ be a solution
to
the contour shrinking conditions relevent toward the identification
of $P$.
The field $J_\vo (y)$ is a dimension 2 field with zero topological
charge.
In the u.v. limit the only such field is the component of the
energy-momentum tensor
$\CT_{zz} (z )$.  Thus to zero-th order in $\la$, $J_\vo \propto \CT_{zz}$.
Any higher order contributions must come from the families
$[ \exp ( in\be (\phi + \phib ) ) ]$ for $n$ an integer, with
dimensions
greater than or equal to $\frac{2n^2 p}{p+1} + k$, for $k$ equal to
some
positive integer.  In order for such a contribution to arise at order
$\la^m$, by dimensional analysis one must have
$$\frac{2m}{p+1} + \frac{2 n^2 p }{p+1} + k = 2 .$$
The only possibility for even $p$ is $k=0, m=n=1$.  This corresponds
to a first order correction in $\la$, which is nothing other than the
order $\la$ correction to $\CT$.  Thus to all orders in $\la$ one has
$$J^\mu_\vo (z,\zb ) ~ \propto ~ \CT^\mu_z  (z, \zb ).$$
The constants $c_p$ in e.g. \IIxxiii, \Vv, can in principle be
computed exactly in the u.v. limit.

Consider now the SG theory at the coupling
\eqn\pprime{ \be = \sqrt{ \frac{2p}{p+ p' } } ~, ~~~~~\Rightarrow q^2 = \exp
\( -2\pi i p' /p \)~, ~~~s = p'/p ~, }
where
$p'$ is a positive, odd integer, relatively
prime with $p$.  By the reasoning in sections 2,3, the
elements $P,\pb$ constructed at $p'=1$ are still well defined
integrals
of motion, but now are viewed as higher spin integrals of motion
with spin $p'$.

\newsec{Applications}
In this section we consider some applications of the
results obtained in the previous sections.  In section 6.1
we review the rerpesentation theory of  quantum affine algebras.
In section 6.2, using the existence of $P$ and $\bar P$ constructed
out of the $Q$'s we show that particles exist
only in deformations of the half-integral spin representations of
$sl(2)$.
Some
of the technical details of this are presented  in appendix B.
In section 6.3 we discuss some generalizations of Witten's
index and the recently discovered index \newind\ in the
context of affine quantum algebras.
We also discuss the notion of `topological twisting'
for these theories.  In section 6.4 we make
a first attempt in formulating a Landau-Ginzburg analog for such
theories.  This subsection, though far from complete, will
discuss some aspects which need to be better understood in order
to derive powerful results analogous to differential equations
satisfied by the new index \cv .

\vfill\eject
\noindent
6.1 ~~{\it Representation Theory}

Recall the representation theory of classical (undeformed) loop
algebras $\hat{g}$. To any finite dimensional representation
$\rho_g$ of $g$, one can associate a finite dimensional
representation  $\hat{\rho}_g \in \rho_g (\nu , \nu^{-1} )$ of
$\hat{g}$,
where $\nu$ is the loop parameter.  Furthermore, every finite
dimensional representation of $\hat{g}$ is isomorphic to tensor
products of
such representations.  These tensor product representations
are irreducible if and only if the loop parameters $\nu_i$ are
distinct.  As was shown in \ref\chari{V. Chari and A. Pressley,
Commun. Math. Phys. 142 (1991) 261.}, when $q$ is not a root of
unity, this structure persists
for affine quantum algebras as well. Let us describe more precisely
these finite dimensional representations at generic $q$.

Let $e,f,h$ satisfy the relations of the finite $\CU_q (sl(2))$
algebra
in the standard presentation:
\eqn\VIi{
[h,e] = 2 \, e ,~~~~~[h,f] = -2\, f ,~~~~~[e,f] =
\frac{q^h - q^{-h}}{q -q^{-1}}. }
Given any representation  of this algebra, a representation
of affine $\uq$ can be
constructed
from the isomorphism:
\eqn\VIii{
Q_+ = a\, \nu \, e\, q^{h/2} , ~~~~~
Q_- = a\, \nu \,  q^{-h/2} f , ~~~~~
\Qb_- = a\, \nu^{-1}  \, f\, q^{h/2} , ~~~~~
\Qb_+  = a\, \nu^{-1}  \,  q^{-h/2} e . }
The $\uq$ relations \IIi\IIii\ are easily verified using \VIi.
The relations \IIi\ are straightforward;  for the Serre relations,
one can show they are equivalent to
\eqn\VIiii{
-[f, e^3 ] + (1 + q^2 + q^{-2} ) e \, [f,e] \, e = 0 , }
which again easily follows from \VIi.
Note that for representations that satisfy $f= e^\dag$, which is
compatible with \VIi, the representations of $\uq$ constructed in
this fashion have the hermiticity properties \IIiv. All finite
dimensional representations depending on a single free loop
parameter can be constructed in this way.  A few low dimensional
examples are given in appendix A.  Other finite dimensional
representations correspond to tensor products
of the above ones, and depend on more than one loop variable.
These tensor product representations are generally irreducible,
unless certain relations are satisfied among the various $\nu_i$.
(See \chari.)

When $q$ is a root of unity, the representation theory is
significantly modified. We make the reasonable assumption that
the above structure of finite dimensional representations of
$\uq$ is still valid, except that one must replace the
representations of the finite $\CU_q (sl(2))$ with the
appropriate ones at a root of 1.  The representation theory of
the finite
$\CU_q (sl(2))$ in this situation is well-studied\ref\pas{V. Pasquier
and H. Saleur, Nucl. Phys. B330 (1990) 523.}\ref\keller{G. Keller,
Lett. Math. Phys. 21 (1991) 273.}\ref\roche{P. Roche and
D. Arnaudon, Lett. Math. Phys. 17 (1989) 295, and
preprint CERN-TH.6416/92.}\lustig.  One finds that due to the fact
that $e^p , f^p $ are central, finite dimensional representations are
at most $p$-dimensional.  The type A representations
$\rho^{(j)}$ are characterized by $e^p = f^p = 0$, and
are deformations of the classical spin $j$
representations of dimension $2j+1$, where $j= 0,1/2,.., (p-1)/2$.
Within this class, one distinguishes between Type II, with
$j\leq p/2 -1$, and Type I, with $j=(p-1)/2$ plus certain
`mixed' representations with $q$-dimension
$[n] = (q^n - q^{-n})/(q-q^{-1} )$ equal to zero. Type B, the
so-called
periodic representations, are $p$ dimensional, and have no classical
analogue.  They are characterized by
$e^p = f^p = 1$, which is also consistent with centrality of
these elements\foot{There is also a $Z_2$ doubling of representations
\chari\ which is not relevent for what we are considering here.}.

In view of the relations \IIxx, it is only the Type
A representations that are relevent for the considerations of this
paper.
However,
one can apply the basic constructions  of our framework to
hypothetical
theories with
Type B periodic representations.
In the periodic representations one would have $Q_\pm^p , \Qb_\pm^p
\neq
0$.
The results of section 3 show that the contours can be shrunk in $\a
Q_\pm^{p-1} (Q_\pm )$.
In any theory with genuine $U(1)$ symmetry generated by $T$, one must
have
\IIxx.  However in other models with no $U(1)$ symmetry but rather
a $Z_p$ symmetry (which occurs in  the algebra $\uq'$), the central
elements
$Q_\pm^p , \Qb_\pm^p$ are neutral, and given the Lorentz spin of
these operators, the following identification is consistent:
\eqn\cyclic{ Q_\pm^p = P , ~~~~~~\Qb_\pm^p = \pb , }
where here $P, \pb$ have Lorentz spin $p'$.
Note that  here $p$ can be even or odd.
This is reminiscent of
the structure that was found in certain perturbations
of $Z_p$ invariant conformal field theories\ref\zf{A. Zamolodchikov,
Sept 1989 preprint, unpublished, V. A. Fateev, Int. J. Mod. Phys. A6
(1991)
2109.}.  However the latter models do not
have the full quantum affine structure (they only have two charges
rather
than four),
but rather a $\CU_q (sl(2))$ restriction of $\uq$  that is rather
a generalization of $N=1$ supersymmetry; moreover these models
have no apparent Type B
periodic representations.
In \ref\pasber{D. Bernard and V. Pasquier, Int. Journ. Mod. Phys. B4
(1990)
913.}\
it is proposed that the  relation \cyclic\ is relevent to the chiral
Potts model of classical statistical mechanics at the $Z_p$ point,
however it is unclear if this structure persists in
the continuum quantum field theory.
Thus we know of no quantum field theory which is a precise
realization
of this possibility.

\def\roj#1{\hat{\rho}^{(#1)}}

\bigskip
\noindent
6.2 ~~{\it Spectral Properties}

We now use the above results to derive some general properties of
quantum field theories with the symmetry $\uq$.
We assume we are given a theory with $\uq$ symmetry, such that
at certain points $q^2 = \exp (- 2\pi i /p )$, and $P,\pb$ are
given by constructions of the previous sections.
Let us further suppose that
the $\uq$ symmetry is realized on asymptotic particle states.  The
basic
finite dimensional representations $\hat{\rho}^{(j)}$
constructed above which depend on a single loop parameter can be
interpreted as single particle representations, where, as discussed
above for $\roj {1/2}$, the loop parameter is $\nu = e^{s\, \th }$,
where
$\th $ is the rapidity of the particle, and $s$ is the Lorentz spin
of
the charges $Q_\pm$.  The tensor product representations are of
course
understood as multiparticle states.

$\uq$ invariant S-matrices for the integrable theories
can be constructed as follows\foot{Subsets of these
S-matrices may be valid for the non-integrable cases
as well.}. Let
$S^{j,j'} (\th_1 - \th_2 )$ denote the 2-particle to
2-particle S-matrix for the scattering of spin $j$ with
spin $j'$ particles.  Since the $\uq$ symmetry acts on
2-particle states via the comultiplication $\De$, the
invariance of the S-matrix is the statement
\eqn\sym{
S^{j,j'} (\th_1 - \th_2 ) \>
\roj j \ot \roj {j'} \( \De (A) \)
{}~ = ~
\roj j \ot \roj {j'} \( \De ' (A) \) \> S^{j,j'} (\th_1 - \th_2 ) ~~~~~
\forall \, A \in \uq , }
where
$\De'$ is the permuted comultiplication, i.e. $\De' = \sigma \, \De$,
where
here $\sigma$ is the permutation operator: $\sigma (u\ot v ) = v \ot u $.
These equations were studied in \jimbo, where it was shown that
solutions are unique up to an overall scalar factor and automatically
satisfy the Yang-Baxter equation.  Solutions can also be
constructed from the fusion
procedure\ref\fusion{P. P.
Kulish,
N. Yu. Reshetikhin, and E. K. Sklyanin, Lett. Math. Phys. 5 (1981)
393.}.   Thus the S-matrices take the form
$S^{j,j'} (\th ) = s_0 (\th ) \, R^{j,j'} (\th )$, where
$R^{j,j'}$ are the R-matrices constructed in \jimbo, and
$s_0 (\th )$ is an overall scalar factor which makes
$S^{j,j'}$ crossing symmetric and unitary.  The minimal
solution for $j=j' = 1/2$ is the known SG S-matrix\foot{See \rbl\
for details of this correspondence.}.  For $j=j'=1$,
the same S-matrix was constucted independently in
\ref\spinone{A. B. Zamolodchikov and V. A. Fateev, Sov. J. Nucl. Phys.
32 (1980) 298.}\ by imposing the Yang-Baxter equation directly.
The advantage of constructing the S-matrices from \sym\ is that
it fixes the dependence of the S-matrices on the physically relevent
parameters, which are encoded in the value of $q$.

One can evaluate $P,\pb$ for each of the finite dimensional
representations of $\uq$.  One may do this by using explicitly
the formulas for $P, \pb$ and the explicit representations
$\roj j$. However the result can be determined purely from
the abstract algebraic properties of $P,\pb$ without knowing
explicit formulas for  $P,\pb$  and $\roj j$.  We henceforth
assume simply that $P, \pb$ exist for any $p$.  We normalize the
constants $c_p$ in e.g. the formulas \IIxxiii, \Vv, by
rescaling the $Q$'s if necessary (which may amount to a
redefinition of $a$ in \IIi ) so that
$$P = \epsilon \( \, \( Q_+ Q_- \)^{p/2} + \( Q_- Q_+ \)^{p/2} + ....
\)$$
where $\epsilon =\pm 1$.  It is not a priori clear which
choice of sign is realized physically (except in the $p=2$ case
where positivity of $P$ implies that $\epsilon = +1$).  We find that
 $P= \epsilon \,  a^p e^\th $ in the representation $\roj {1/2} $.
Then one can show that for any $p$, in the
representations $\roj j$ one has
\eqn\VIiv{\eqalign{
P = \epsilon \, a^p \, e^\th ~ (-1)^{j-1/2} ,  ~~~~~~
\pb = \epsilon \, a^p \, e^{-\th}  ~ (-1)^{j-1/2} ,  ~~~~~~ {\rm if}
{}~~
&j =   1/2, 3/2, .., (p-1)/2  \cr
P = \pb = 0 ~~~~~~~~~~~~~~~~~~~~~~~~~~
{\rm if} ~~ &j = 0,1,..,(p-2)/2 . \cr }}
The proof of these results relies on certain elementary aspects
of the fusion procedure and is given in appendix B.  The basic
idea behind the proof is that due to its trivial comultiplication
$P$ acts very simply on tensor product representations, thus one
can deduce its value in any representation obtained by
decomposing tensor products.
One may verify the above results explicitly for the representations
listed in appendix A using the explicit form of $P,\bar P$ given in
the previous sections for $p=2,4,6$.

The result \VIiv\ has interesting implications, which we now discuss.
The formula \VIiv\ implies that integer spin $j$ (including $j=0$
singlet)
representations
of massive particles are excluded, since for any $a$ these
representations have zero energy.  If anything, they can only
correspond to degenerate vacuum representations.  If this happens
this means that affine quantum symmetry is spontaneously broken
as the affine quantum charges do not annihilate the vacuum.
One also sees that in a given model, with fixed $\epsilon$, the
energy alternates in sign for the half-integer spin representations,
thus they are not all allowed if one requires positivity of the
energy.    In a given theory, one can have at most either
the spin $j$ representations with $2j = 1 ~{\rm mod}~4$,
or $2j = 3 ~{\rm mod} ~4$ depending on whether $\epsilon =+1$
or $-1$.
The mass $m$ of the particles in these representations is given
by
\eqn\mass{
m^2 = P \pb = a^{2p} , }
and is independent of the spin $j$.

One may ask whether all of the latter half-integral
spin representations are allowed in a single theory.
Consider an integrable theory where the S-matrix bootstrap
axioms hold.
If a theory contains spin $j'$ particles in addition
to spin $j$,  and if the S-matrix $S^{j,j}$ contains poles
for intermediate spin $j'$ particles, then the S-matrix
$S^{j,j'}$ can be reconstructed from $S^{j,j}$ using the
closure axioms of the bootstrap.
The fusion procedure, which describes how tensor products of
$\uq$ representations decompose, can be understood as this closure
property of the S-matrix.  For half-integral $j$, since
$\roj j \ot \roj j$ can only be decomposed into integral
spin $j'$ representations, this implies it is impossible to find
poles corresponding to other half-integral spin $j'$ particles
in $S^{jj}$.\foot{The $N=2$ case is exceptional in this regard, as
any shift $T \to T + \al$, where $\al$ is an arbitrary constant, still forms
a doublet representation of the $N=2$ superalgebra with a different
value of $a$.}\
Thus it appears likely that one can
only have one-particle states in a  single representation
$\roj j$ of $\uq$, and $j$ must be half integral.

Let us compare the above selection rules with the spectrum of
the known models with $\uq$ symmetry.
It can
be inferred from the papers \rbl\ that all of the known models
can be described universally as anisotropic perturbations
of the level $k$ $su(2)$ Wess-Zumino-Witten (WZW) models with the
action
\eqn\wzw{
S = S_{WZW}^{(k)} ~ + ~ \la \int d^2 x
\( J^+_\mu J^-_\mu  + J^-_\mu J^+_\mu + g \, J^3_\mu J^3_\mu \)
, }
where $J^a_\mu (x)$ are the local level $k$ $su(2)$ currents of
the WZW theory, and $g$ is the anisotropy parameter.
When $g =1$, the diagonal $su(2)$ is unbroken. When
$g \neq 1$, this theory has a $\uq$ symmetry for any $k$, where
$q$ is related to $g$ in a precise way that can be derived by
bosonizing  the WZW theory with a $Z_k$ parafermion
and a free boson and relating the model to the $k$-th fractional
supersymmetric SG model\foot{When $g = 1$, the free boson is
compactified at a radius $r= 1/\sqrt{2k}$, and deforming $g$
away
from $1$ simply changes $r$.  The resulting value of $q$ is
$q= -\exp ( -i\pi ( 1/(2k^2 r^2 ) - 1/k ) ) $ \rbl.}.
The SG model occurs at $k=1$.
All of these models have points in the coupling space where
$q$ is as given in \IIxvi.
The higher $k$ theories would be
likely candidates for realizations of the
higher spin representations since they
contain chiral primary fields of spin $j \leq k/2$.
Certainly, the above results imply there cannot be any integer
spin $j$ representations in the spectrum.
In the case of the
SG theory this confirms the well-known result that for values
of the coupling \eix\ there are no singlet bound states (breathers).
Actually,
it was found that for all $k$ the particles in the spectrum still transform in
the spin $j=1/2$ representation, the S-matrix being
the tensor product of the SG S-matrix and an RSOS S-matrix.

The field theories, if they exist,  which correspond to the scattering given
by $S^{j,j}$ for $j> 1/2$ have not yet been found.  At least for integer spin
$j$ our results rule out the existence of such theories with
$q^2 = \exp (- 2\pi i / p )$ and $P, \pb$ given by the above constructions.
However it is possible that some theories do exist which exhibit
the scattering $S^{jj}, ~ j\in \Zmath$, if the expressions we have
constructed for $P,\pb$ cannot be identified the momentum operators but
rather are identically zero. This could be the case for instance if the
theory is such that the Lorentz spin of the charges $Q_\pm$ is not
$1/p$.  As an illustration of this statement consider again the
SG theory.  When $\be^2 < 1$, singlet breather states are known to
appear in the spectrum. From eq. \pprime\ one sees that this requires
$p<p'$.  Thus, precisely in the region of the coupling where singlets
appear, the elements $P, \pb$ we have constructed have Lorentz spin
$p' > 1$, and are not the true momentum operators;  integer $sl(2)$ spin
states are now allowed and indeed breathers found in the spectrum.

The relation \mass\
shows that the basic one-particle states in each representation satisfy
the Bogomolnyi bound just as in the $p=2$ case.
In  specific models one can relate $a$ and thus the mass to certain
topological
characteristics of the theory.
For example in the SG model,  from \xtop\ we have
\eqn\VIv{
\frac{\la^2 \gamma^4}{4 \pi^2 } \> |\Delta X |^2
= a^4 \bigl| \frac{1-q^{2T}}{1-q^2} \bigr|^2  , }
where $\Delta X = X(\infty) - X(-\infty )$.
In the spin $j=1/2$ representation, the right hand side of
\VIv\ is just $a^4$, thus one obtains
\eqn\VIvi{
m^2 =  \( \frac{\la \gamma^2 }{2\pi } \> |\Delta X | \)^p  ~~~~~~~~(j=1/2). }
(A choice of mass units is implied in the above to avoid
an overall constant ambiguity).

\def\bt{\tilde{\beta}}

\bigskip
\noindent
6.3 ~~{\it Generalizations of Supersymmetric Indices}

An important tool in the study of supersymmetric theories is
Witten's index $Tr (-1)^F $, where $F$ is fermion number.  Let
$\CH$ denote the Hilbert space and consider
\eqn\VIvii{
I = Tr_\CH ~ (-1)^{T/2} \> e^{-\tilde{\beta} H } , }
where $H$ is the Hamiltonian and $\bt$ is a constant (inverse
temperature).  At the supersymmetric point $(p=2)$, fermion
number $F$ is defined to be $\pm 1$ for the charges $Q_\pm$, thus
$F= T/2$, and \VIvii\ is the Witten index.  For higher $p$,
the index $I$ shares many of the desirable properties at the
$p=2$ point.  First it is easy to see that all of the allowed
half-integer spin $j$ one-particle representations of $\uq$
contribute zero to $I$.   Note that the
odd-dimensional integer spin $j$ representations, if they
exist in the theory, are not projected out by $(-1)^{(T/2)}$,
but since they have $H=0$, they also do not introduce
any $\bt$ dependency!   So it is precisely the miracle
of having no (massive) odd dimensional representation in the theory
which makes this index independent of $\bt$ for one particle states.

The above results indicate that it is possible to show that
$I$ is independent of $\bt$ for one particle states by using the
algebraic
properties described in the previous sections.  Namely,
\eqn\VIviii{
\d_{\bt} I = - Tr_\CH (-1)^{T/2} (P + \pb ) \> e^{-\bt H }  = 0 .}
To prove this result using the $\uq$ algebraic properties, one
can insert the expressions for $P,\pb$ in terms of $Q_\pm , \Qb_\pm$.
Since $Q_\pm , \Qb_\pm$ commute with $H$, they can be cycled in the
trace:
\eqn\beep{ Tr (-1)^{T/2} A\, Q_\pm \, e^{-\bt H} =
 - Tr (-1)^{T/2} Q_\pm \, A\, e^{-\be H} , ~~~~~~~~~A\in \uq . }
 For $p=2,4$, using the expressions \IIxxiii\ one sees that the terms
 in $P$ and $\pb$ cancel in pairs in the expression \VIviii.
For higher $p$ we have not proven \VIviii\ in  this algebraic
fashion, but the arguments of the last paragraph indicate it must
be possible.

Now we have to consider whether this index
is independent of $\bt$ also for multi-particle
states.  One would {\it naively} expect
that any tensor products of
such representations contribute zero. This expectation is
based on assuming that the degeneracies of multiparticle
states are the same in a given representation of
affine quantum group.  However, it is well known that
such expectations are generally invalid
and that  the degeneracy of states in a given
multiplet will not generally be the same.  This was indeed
the reason that the new index defined in \newind\ is
$\bt$ dependent, in a non-trivial way.  So a priori it could be that the
$I$ defined above depends on $\bt$ from multi-particle
contributions.  In the context of Sine Gordon model
this has been recently checked \ref\fein{P. Fendley and
K. Intriligator, private communication.}\ and it has
been found that $I$ {\it does} pick up a two particle
contribution\foot{
This is different from the $p=2$ case where Witten's index
is independent of $\bt$.}.  In fact this result
is not surprising, because in the conformal limit we
do not have any natural index, and all twisting
(changing of the boundary conditions) of the
$c=1$ model give rise to partition functions which depend
on $\bt$ (or the moduli of torus).  It would be very interesting
to find a simple differential equation characterizing
the $\bt$ dependence of this object.

We can also go on and define another `index' in analogy
with \newind  .  The object to consider would be
$$Q=Tr (-1)^{T/2} T\ e^{-\tilde \beta H}$$
It is easy to see that just as in \newind\ this
index receives contributions from one particle
states and again {\it naively} receives no contribution from
multi-particle states.  However there is an anomaly
and one can compute for example the two particle contribution
to $Q$ which is non-vanishing.  It would be very interesting
to develop the machinary just as in \cv\ and derive
differential equations which characterize $Q$ for all even $p$.

\def\ar{{\rm ad}^R \, }
The first step toward a deeper understanding of the higher $p$
algebras is to find the analog of `topological twisting'
which exists in the $p=2$ case.  In the $p=2$ case one considers
coupling the theory to a background $U(1)$ gauge field, which
is coupled to the $T/2$ current, and is set equal to ${1\over 2}$
of the spin connection \EW \ref\ttwi{T. Eguchi and S.K. Yang, Mod.
Phys. Lett.
A5 (1990) 1693\semi
C. Vafa, Mod. Phys. Lett. A6 (1991) 337
\semi R. Dijkgraaf, E. Verlinde, and H. Verlinde, Nucl. Phys. B352
(1991) 59.}.
  In this way
the spin of fields change according to
$$s \rightarrow s-{\tilde{q}\over 2}$$
where $\tilde{q}$ is the charge $T/2$ of the state.
In this way $Q_+$ and $\bar Q_-$ become scalars and
$Q_-$ and $\bar Q_+$ become spin $\pm 1$.  Moreover the
theory has an energy momentum tensor which can be written
as (anti)commutators of $Q_+ +\bar Q_-$ with some operator. In the
conformal case this means that the twisted theory has $c=0$.

The natural generalization of this for higher $p$ is to set the
gauge field which couples to $T/2$ to be ${1\over p}$ of the
spin connection.  In this way the spins shift according to
\eqn\sch{s \rightarrow s-{\tilde{q}\over p} .}
Alternatively this corresponds to shifting
$L \rightarrow L - T/2p$.
In this way $Q_+$ and $\bar Q_-$ become scalars and $Q_-$
and $\bar Q_+$ have spin $\pm 2/p$.  Note that $Q_+$ and $\bar Q_-$,
which
generate the quantum group algebra $\uq^{(0)}$ as a subalgebra
of $\uq$ is the true symmetry in the twisted version\foot{In the
mathematical
terminology this twist corresponds to going from the principal
to the homogeneous gradation.}.  This suggests
a major difference with the $p=2$ case, in that both $Q_+$ and $Q_-$
cannot have integral spins in any given sector (as they have
spins which are different mod integers in the twisted version). Put
differently, if we quantize on a circle the operators $Q_+$ and
$Q_-$ are not defined on all states in a Hilbert space.  In
particular if $Q_+$ is defined on the subsector $\cal H_+$ and acts
by
$\cal H_+ \rightarrow
\cal H_-$ then $Q_-$ is defined as the adjoint operator
acting as $\cal H_- \rightarrow
\cal H_+$.
This structure is reminiscent of Felder's construction
of the minimal models from $c=1$ theory\ref\felder{G. Felder,
Nucl. Phys. B317 (1989) 215.}, and shows that there exists
a massive analog of Felder's cohomology, based on \IIxx.
Indeed the above
topological twisting applied to the $c=1$ model
is nothing but changing the energy momentum
tensor by background charge, which shifts the central charge
to $c=1-{6\over p(p+1)}$.  Note that unlike the $p=2$ case
the twisted theory has $c\not= 0$.
This is mirrored by the fact that the $P$ that
we obtained cannot be written solely as a
(generalized) $Q_+$ commutator.  We also
need $Q_-$ commutators.  At any rate in the conformal
case the `topological states' are to be identified with the
non-trivial elements of Felder cohomology which is simply
the Hilbert space of minimal models.  So we see
a major difference with the $p=2$ case, in that there are
apparently infinitely many topological states, as opposed to the
$p=2$ case where one typically obtains a finite number of states.

There exists another meaningful trace which is motivated
by the fact that $\uq^{(0)}$ is a symmetry of the twisted
theory.  To introduce this let us first note the
following: The adjoint actions used above
are left actions.  Using the same notation as in \IIvi\ define a
right
action as
$$ {\rm ad}^R A (B) = \sum_i \, S' (b_i ) \, B \, a_i , $$
where $S'$ is the skew antipode defined to satisfy
$m(S' \ot 1 ) \De' = \ep $.
This satisfies
$$\ar A \, \ar B (C) = \ar BA (C) . $$
For $\uq$ these take the form
$$\ar Q_\pm (B) = - q^{\mp T_B } \, \a Q_\pm (B) ~~~~~~~
\ar \Qb_\pm (B) = - q^{\pm T_B } \, \a \Qb_\pm (B) .$$
Now let us consider the following `index'
$$I' = Tr_{\CH} \, q^{-T} \, e^{-\bt H}  . $$
This trace satisfies
$$ Tr q^{-T} \, A \a Y (B) = Tr q^{-T} \a^R Y(A) B , ~~~~~\forall
 ~ Y \in \uq^{(0)} . $$
Since this trace is only invariant under the $\uq^{(0)}$ subalgebra,
$I'$ is not independent of $\bt$. However since the symmetry
of the twisted theory is nothing but $\uq^{(0)}$, this is presumably
an important object to consider for the topologically twisted
theory.  The fact that it is not independent of $\bt$ may be
related to the fact that there are infinitely many topological
states and they appear at all energy levels, just as in the minimal
model case.
 An interesting property of $I'$ is that it projects out all of the
Type AI representations\foot{One
can show that
$$
Tr e^{2\pi i L } A \, \a Y (B) = Tr e^{2\pi i L} \ar Y(A) \, B , ~~~~~~
\forall ~ Y \in \uq , $$
where
$L$ is the Lorentz boost operator.  The virtue of the
trace $Tr e^{2\pi i L}$ is these invariance properties under
adjoint action for the full quantum affine algebra $\uq$. However,
one cannot use this trace  to define an index
since the operator $e^{2\pi i L}$ is
not well-defined on eigenstates of $H$, due to the fact that
$L$ does not commute with $H$.
This was to be expected as there is no twisting
which makes all Q's have integral spins at the same time. However
this trace is well-defined on the space corresponding to the action
of fields at a single space-time point on the vacuum.}.
\bigskip
\noindent
6.4 ~~{\it Generalized Superspace Landau-Ginzburg Reformulation }

In the $p=2$ case, i.e. the standard $N=2$ supersymmetric theories,
a deeper understanding of the structure of the theory in which
the supersymmetry is manifest comes from a superspace formulation.
In addition to the spatial coordinates one introduces anti-commuting
odd coordinates $\theta^{\pm}$ (and the right-moving counterparts).
The fact that they anti-commute is modeled after
$Q_\pm ^2= 0$ and $\{ Q_+,Q_-\} =0$
(on the $P=0$ subspace).  In the affine quantum case with
$p>2$ in order to understand the appearance of $\uq\ $ symmetry
in a natural way one would still have to introduce two extra
coordinates
$\theta_\pm$ (and the right-moving counterparts).  However
now the commutation properties of them are a little more
complicated.  Modeled after $Q_\pm^p =0$ one should impose
$\theta_\pm^p =0$.  The complication arises when
we consider assigning commutation properties {\it between}
$\theta_+$ and $\theta_-$.  In this case the  natural
relation we can impose is modeled after the Serre relation \IIii :
$$\theta_\pm^3\theta_\mp- (1+q^2+q^{-2})\theta_\pm^2\theta_\mp
\theta_\pm +(1+q^2+q^{-2})\theta_\pm \theta_\mp \theta_\pm^2
-\theta_\mp \theta_\pm^3 =0.$$
To this we should also add the constraint $P=0$ in determining
the commutation properties of $\theta$'s (just as in the $p=2$ case).
Apart from these, there is no other natural commutation relations
among
these $\theta$'s.
{}From this it seems clear that the superspace formulation is
going to be much more involved than in the $p=2$ case, as
it seems that the fields do not have a finite expansion
when expanded in superspace.   In other words, we will
end up with infinitely many {\it auxiliary fields}.  It would
be very challenging and interesting to unravel the structure
of the corresponding field theory; this  project  is  beyond
 the scope of the present paper.

At this point one has the option of postponing a discussion
of a Landau-Ginzburg theory in a superspace until such
a formulation has been studied in more detail.  However, assuming
that the superspace can be made sense out of, and that the
basic structure is not too drastically different from the $p=2$
case, we can make some natural extrapolations to higher $p$ and
guess at least some aspects of such LG theories.
Furthermore, one may develop many aspects of the theory without
reference to superspace, i.e. in `components'.
This is what
we will presently do in the context of Sine-Gordon model.
The following results are not
completely rigorous, but are noteworthy enough for us to present
them at this stage.

First let us recall some basic facts about the $p=2$ case
\ref\landgins{E. Martinec, Phys. Lett. 217B (1989) 431\semi C. Vafa
and N.P. Warner, Phys. Lett. 218B (1989) 51\semi
S. Cecotti,L. Girardello and A. Pasquinucci, Int. J.
Mod. Phys. A6 (1991) 2427.}.  In this case one considers
`chiral'\foot{This notion of chirality
is not to be confused with the  usage of the term in previous sections
of this paper. See footnote
8.} fields $X$ (which commute with $Q_+$) and form two
types of terms in the action:  The $D$--term which
involves integration over all super-coordinates and involves
a function of both $X$ and its complex conjugate $K(X,X^*)$.  Then
there is also an $F$--term, which involves integration over
half the superspace of a function $W(X)$ holomorphic in $X$ (and
its conjugate).  $W$ is known as the superpotential.
The action takes the form
$$S=\int d^2z d^4\theta \ K(X,X^*)+\( \int d^2z d\theta_- d\bar{\theta}_+
\
W(X)+h.c. \)$$
which gives the equation of motion
\eqn\emot{D^-\bar D^+ ({\partial K \over \partial X})+{\partial
W\over
\partial X}=0}
A basic fact
is that under topological twisting $X$ has $0$ left (and right)
dimension
which naturally leads to the concept of chiral rings.

Now we try to mimic this structure for the $p>2$ case of Sine-Gordon
theory.  From the
relations \xtop,   one sees that the $\uq$ algebra closes on
topological charges for the fields $X, X^*$.  Since in the
topological setting $\uq^{(0)}$ is a good symmetry, based on
analogy with $p=2$ one would naturally take $X$ as the analog
of (anti-)chiral field. This identification is strengthened
by the fact that in the topologically (or
anti-topologically) twisted version $X^*$ (or $X$) will have zero
dimension, as follows from \sch .  Thus
we take $X$ as the basic LG field.

We first rewrite the SG potential $\cos (\be \Phi )$ in terms of the
fields $X, X^*$ and the $\uq$ charges.  Let $2 :\cos (\be \Phi ): = V +
V^*$,
where $V= :\exp ( i\be \Phi ):$.
At the values of the coupling \eix, $X= \exp (-i
\sqrt{\frac{2}{p(p+1)}}
\Phi )$.
The fields $V, V^*$ have (untwisted)
scaling dimension $2p/(p+1)$ and $X,X^*$
have
dimension $2/(p(p+1))$.  The coupling $\la$ has dimension $2/(p+1)$.
One has the basic formulas
\eqna\IIIi
$$\eqalignno{
\a \q- \qb+ (V) &= \gamma^2 \> \d_z \d_\zb X , ~~~~~\a \q+ \qb- (V^*
)
= \gamma^2 \> \d_z \d_\zb X^*  &\IIIi a\cr
\a \q+ \qb- (V) &= \a \q- \qb+ (V^* ) = 0
. &\IIIi b \cr}$$
The above equations are proven by breaking up $X$ and $V$ into their
quasi-chiral components and using conformal operator product
expansions.
 In fact, the chiral part of \IIIi{} is what ensures that the $\uq$
charges are conserved (see \eres ), so these relations are intimately
tied to
the existence of the $\uq$ symmetry itself.

\def\qt{{\tilde{Q}}}
\def\qbt{{ \tilde{\bar{Q}} }}

The expressions for $P, \pb$  can now be used to express $V$ and
$V^*$ as
an action of elements of $\uq$ on the fields $X, X^*$.  Define
$\tilde{Q}_\pm , \tilde{\bar{Q}}_\pm \in \CU_q (\hat{sl(2)} )$ by
the formulas
\eqn\IIIii{
P = \q+ \qt_- + \q- \qt_+ , ~~~~~\pb = \qb+ \qbt_- + \qb- \qbt_+ . }
The $\qt_\pm , \qbt_\pm$ can be read off from the explicit
expressions for $P, \pb$.  For example
for $p=4$, $\qt_- = c_4 (\q- \q+ \q- - Q^2_- \q+ - \q+ Q^2_- )$.
Let us further suppose that the analog of `chirality' conditions hold:
\eqn\anni{
\a \qt_- (X) = \a \qbt_+ (X) = \qt_+ (X^*) = \qbt_- (X^*) = 0.}
The relations \anni\ can be established purely algebraically. Consider
the first relation at $p=4$.  From \xtop\ one has
$$
\a \qt_- (\De (X) ) \propto \a \qt_- \a Q_- \( \Qb_+ \) .$$
{}From the Serre relations one finds $\qt_- Q_- = - Q_-^3 Q_+ $,
thus
$$
\a \qt_- (\De (X) ) \propto \a Q^3_- \a Q_+ \( \Qb_+ \) = 0, $$
where we have used \IIix.
We have also verified directly in the field theory using contour
shrinking conditions and scaling arguments
that the chirality conditions \anni\ are
valid in some simple cases.
For $p=2$ this is easily proven.
The contour shrinking conditions which are generally
needed to prove \anni\ are of
precisely
the same kind as in section 3, due to the following braiding
relations
\eqn\xbraid{
J_\pm (x) \, \chi (y)
 ~=~ q^{\mp 2} \chi (y) \, J_\pm (x), ~~~~~
J_\pm (x) \, \chi^* (y)
 ~=~ q^{\pm 2} \chi^* (y) \, J_\pm (x)
 ~~~~~x<y. }
This means that $\chi , \chi^*$ behave just like $J_-$ and $J_+$
respectively in the braiding relations.
Thus to prove for example that $\a \tilde{Q}_- (X) = 0$ when $p=4$,
one must show that
$$ M\, \( v(-+--) - v(--+-) - v(+---) \) = 0 . $$
This is easily verified; in fact $\a \tilde{Q}_- (Q_- ) = 0$ is just
a different expression for the Serre relation, which is only possible
since $q^4 = -1$.
Assuming \anni,  \IIIi{} and \IIIii\ imply
\eqn\IIIiii{
V = \gamma^2 \>  \a \qt_+ \qbt_- \( X \) , ~~~~~~~
V^* = \gamma^2 \> \a \qt_- \qbt_+ (X^* ) . }

Define powers
$X^n (0)$ of the field $X$
by the (regularized) operator product expansions:  $X^2 (0) = \lim_{\ep \to 0}
X(\ep) X(0) $, $X^3 (0) = \lim_{\ep \to 0} X(\ep) X^2 (0)$, etc. The
operator product expansion implies
\eqn\IIIiv{
X^p (0) = \lim_{\ep \to 0}
\ep^{\frac{2(p-1)}{(p+1)}} ~ V^* (0). }

Let us {\it define} the integrals $\int d \th_\pm $, $\int d \bar{\th}_\pm$
by the formulas
\eqn\intth{
\int d\th_\pm \, W \equiv \a \qt_\mp ( W) , ~~~~~
\int d \bar{\th}_\pm W \equiv \a \qbt_\mp (W) . }
Now we propose, at the conformal point, the action of the form
$$\int d^2z d^4\theta \ K(X,X^*) + \int d^2z d\theta_-
d\bar{\th}_+ \ W(X)+h.c.$$
where to leading order we take $K=X X^*$ and $W=X^{p+1}/(p+1)$.
Indeed the equations of motion \emot , with this choice of $K$ and
$W$
becomes
$$ \a \qt_- \qbt_+ (X^* )+X^p=0$$
which (up to normalization questions) follows from \IIIiv\
and \IIIiii !  Next we add perturbation $V+V^*$ which takes
us aways from the conformal point.  In view of \IIIiii\
we can represent this perturbation by the addition of $X$
to the superpotential, and so we end up with the action
\eqn\IIIvii{
S = \int d^2z d^4\theta \> X X^*  + \( \int d^2 zd^2\theta \( \inv{p+1}
X^{p+1} - \hat{\la} X \)
+ h.c. \) }
(where $\hat{\la}$ is a rescaled $\la$).
In the $p=2$ case the superpotential would suggest a `chiral ring' of
the
form $X^p=\hat{\la}$.  Is this the case for us?\foot{The definition
of such a ring for general $p$ is bound to be more subtle than in
the $N=2$ case, and is beyond the scope of the present paper.}
Consider $ <X^p (0) > $ in conformal perturbation theory. One has
\eqn\IIIv{
< X^p (0) > = <X^p (0)>_{\la = 0} + \frac{\la}{2\pi i} \int d^2 w
< V(w,\bar{w} ) X^p (0) > + \cdots . }
Clearly $<X^p (0) > = 0 $ in the conformal field theory at $\la = 0$.
The only correction is to order $\la$ and it is finite. Evaluating
this
using conformal operator products,
and  regularizing the integral over $d^2 w$, one finds
$< X^p (0) > = - \inv{2} \( \frac{p+1}{p-1} \) \la $.  This
is a further confirmation of the existence of the above
LG formulation of Sine-Gordon theory.

The action \IIIvii\ is manifestly $\uq$ invariant.  To
see this, consider e.g. a variation of $F$--term generated by $\q-$.
One has
$\delta F = \int d^2 z \a \q- \qt_+ \qbt_- (W) = 0$, since by \IIIii\
the integrand is a total derivative. (The variation of the h.c. term
is
zero by \IIIi{b}.)  A similar argument applies to the $D$--term.

The above reformulation of the SG theory has the following
interpretation. The
original $\cos (\be \Phi )$ potential has an infinite number of
minima occuring
at $\Phi = 2\pi n/\be$.  The SG solitons are kinks that interpolate
between
two neighboring pairs of vacua. In terms of the field $X$, from \IIx\
one finds
that the minima in $X$ occur at $X= (q^2)^n $.  But these are
precisely the
solutions of the equations of motion $X^p = \hat{\la} $ which follow
from the action \IIIvii, for $n=0,1,\cdots,p-1$. Thus one may
consider the
theory \IIIvii\ as equivalent to a version of the SG theory where one
limits
the number of minima of the $\cos(\be \Phi)$ potential to $p$ in
number,
i.e. one identifies $\Phi \equiv \Phi + \frac{2\pi}{\be} (p-1)$.
This type of SG theory was recently discussed in \ref\rklas{T.
Klassen
and E. Melzer, `Kinks in Finite Volume',
to appear in Nucl. Phys. B.}.

\newsec{Conclusions}

Based on similarities between $N=2$ supersymmetric algebras and a
special
point of the $\uq$  quantum affine $sl(2)$ algebra, we considered
finding the structures present in the $N=2$ theories
in the  quantum affine  algebras. This led us to discovering new
central elements in these algebras when $q^2$ is an even root of
unity
which we identify with $P,\bar P$.
The existence of such elements required solutions
to overdetermined systems of equations.  We found these   explicitly
for a few non-trivial cases, and indicated how in principle one
could find solutions for all cases.  Although we did not prove
the existence of solutions for all cases, the very fact that in the
first few non-trivial cases that we checked there were unique
solutions, suggests strongly that this is going to be the case
in general.  It would be very interesting to prove this in full
generality and find a general expression for $P,\bar P$ in terms
of $Q$'s.

Assuming the existence of $P,\bar P$ in general, we showed
that representations of integral $sl(2)$ spin cannot be
realized physically, as they will have $P=\bar P=0$.
This was also explicity checked in the cases where we had
expressions for $P$ and $\bar P$.
It would be interesting to find examples of theories
with affine quantum $sl(2)$ symmetry
where particles in higher half-integral spin (other than spin $1/2$)
representations appear.  It would also be interesting
to see what physical implications the absence of integral spin
representations has.

Motivated by similarities with the $p=2$ case  we considered topological
twisting for $p>2$.  In the conformal limit with $c=1$ this is
precisely
Felder's construction of minimal models.  However one obtains
infinitely many `topological states' corresponding to
the Felder cohomology.  It would be interesting to understand
how one could use this structure more effectively, as one
does with ordinary topological theories with $p=2$.
This may lead to a different point of view and perhaps a deeper
understanding of the restricted sine-Gordon theories, which
describe integrable perturbations of the $c<1$ minimal models\ref\lec{A.
LeClair, Phys. Lett. 230B (1989) 103; N. Reshetikhin and F. Smirnov,
Commun. Math. Phys. 131 (1990) 157; D. Bernard and A. LeClair,
Nucl. Phys. B340 (1990) 721.}\rfl.

Related to the above issue, we considered the question of
superspace for higher $p$ and found that it
is bound to be rather subtle.  In particular superfields will contain
infinitely many auxiliary fields.  This would
be very exciting and challenging to develop in a field theory set up.
Assuming this can be done, we proposed a Landau-Ginzburg
like theory for Sine-Gordon theory at even roots of unity.
It would be of great interest to understand the structure
of such Landau-Ginzburg theories.  This may open the
door to obtaining a deeper understanding to a class
of 2d quantum field theories with  quantum affine symmetry.
It should also be interesting to generalize
the above results from $sl(2)$ to other groups.

\bigskip\bigskip
\centerline{{\bf Acknowlegements}}

We would like to thank D. Arnaudon, D. Bernard,
S. Cecotti, G. Felder, P. Fendley, K. Intriligator,
V. Kac, T. Klassen,  A. Lesniewski, and A. Zamolodchikov
for valuable discussions. We are
also
very thankful to G. Ricciardi for assistance in using {\it
Mathematica}.
 The work of A.L. is supported in part by NSF and by an Alfred P.
Sloan Foundation Fellowship.  The work of C.V. is supported in part
by
Packard fellowship and the  NSF grants
PHY-87-14654 and PHY-89-57162.

\appendix{A}{Examples of low-dimensional representations}

Here we provide the spin $j=1,3/2,2,5/2$ representations of $\uq$,
in the following presentation.
\eqn\ai{\eqalign{
T &= {\rm diag}(2j, 2j-1, \ldots , -2j ) \cr
Q_+ &= a \, \nu \, E , ~~~~~ \Qb_- = a\,  \nu^{-1} \, F ,~~~~~
Q_- = a\,  \nu \, E^\dag ,~~~~~\Qb_+ = a\,  \nu^{-1} \, F^\dag  . \cr
}}
with $F=E^t $.  In all cases, $E $ is an upper off-diagonal matrix.
Let  $E_{rs} , ~r,s = 1,..,2j+1$ denote the matrix entries of $E$.
Below we list the non-zero entries $\{ E_{12}, E_{23},...,
E_{2j,2j+1} \}$.

\eqn\aii{\eqalign{
j=1: ~~~~~ &\{ \sqrt{1+q^2} , \sqrt{1+q^{-2}} \} \cr
j=3/2: ~~~~~ & \{
{\sqrt{1 + {q^2} + {q^4}}},
   {\sqrt{2 + {q^{-2}} + {q^2}}} ,
   {\sqrt{1 + {q^{-4}} + {q^{-2}}}}  \} \cr
j=2: ~~~~~ & \{
{\sqrt{1 + {q^2} + {q^4} + {q^6}}}
   , {\sqrt{2 + {q^{-2}} + 2 {q^2} + {q^4}}}, \cr
    &~~~~ {\sqrt{2 + {q^{-4}} + {2 {{q^{-2}}}} + {q^2}}},
   {\sqrt{1 + {q^{-6}} + {q^{-4}} + {q^{-2}}}},
   \}
\cr
j=5/2: ~~~~~ &\{
{\sqrt{1 + {q^2} + {q^4} + {q^6} + {q^8}}},
    {{\left( q^{-1}  + {q^1} \right)  {\sqrt{1 + {q^4}}}}},
   1 + {q^{-2}} + {q^2}, \cr
   & ~~~~
    {{\left( q^{-3}  + {q^{-1}} \right)  {\sqrt{1 + {q^4}}} }},
   {\sqrt{1 + {q^{-8}} + {q^{-6}} + {q^{-4}} +
        {q^{-2}}}}   \}
\cr }}

\appendix{B}{Proof of Spectral Properties}

Here we prove the result \VIiv.  The aspects of the fusion procedure
we will need are the following.  Consider the tensor product
of two representations of $\uq$, $\roj j (\nu_1) \ot \roj {j'} (\nu_2
)$.
If $\nu_1$ and $\nu_2$ are related in a precise way, then this
representation
is reducible\chari.  Let $|j,m>, m\in \{ -j, -j+1,...j \} $ denote the
vectors in the representation $\roj j$ and consider e.g. the
conditions
for $\roj {1/2} (\nu_1 ) \ot \roj j (\nu_2) $ to be reducible
and contain a spin $j+1/2$ representation.
This requires that
\eqn\bi{
Q_- \( |1/2,1/2>\ot |j,j> \) = K_{j+1/2} \Qb_-
\( |1/2,1/2> \ot |j ,j> \) , }
where
$K_{j+1/2} $ is some constant of proportionality.
Using $\De (Q_- , \Qb_- )$ one can compute the action of
$Q_- , \Qb_-$ on the tensor product, and one finds
$$ \nu_1^2 = \nu_2^2 q^{-2} = K_{j+1/2} . $$

Now consider more specifically the case of $\roj {1/2} \ot
\roj {1/2}$. On this tensor product space,
$P = \ep \, a^p (\nu_1^p + \nu_2^p ) = 0$ since $q^p = -1$.  This
shows that $P=0$ on $\roj 1$.  Since the repeated tensor product
of $\roj 1$ with itself yields higher dimensional integer
representations,
this shows that $P=0$ on all integer spin representations.

Next consider reducing $\roj {1/2} (\nu_1 ) \ot
\roj {j-1/2} (\nu_2 )$ for $j$ half integer to obtain the
representation $\roj {j}$.   One has that
$P=\ep \,  a^p \nu_1^p $, where $\nu_1^2 = K_j$.  The constant
$K_j$ can be computed from the definition \bi\ and the defining
relations of $\uq$.  We describe representations using the notation
of appendix A.  In the representation $\roj j (\nu )$ one has that
$\roj j (\Qb_- ) /\nu $ is  a $2j+1$ dimensional matrix with
only upper off diagonal entries $F_{rs}$, (with $r= s+1$), and
$\roj j (Q_- ) \nu^{-1} = F^*$.  One has then $K_j =
F_{21}/F^*_{21}$.
{}From the relation $\IIi$, one finds that
$F_{21} = \( ({1-q^{4j}}) /({1-q^2}) \)^{1/2}  $.
Thus $K_j = q^{2j-1}$, and this proves \VIiv.

\listrefs
\end